\documentclass[amsmath,amssymb]{revtex4}
\usepackage[english]{babel}
\usepackage{natbib}
\usepackage{multirow}

\usepackage{rotating} 
\usepackage{longtable}
\usepackage{color}

\begin{document}

\title{ Properties of Reactive Oxygen Species by Quantum Monte Carlo}

\author{Andrea Zen}
\email{andrea.zen@uniroma1.it}
\affiliation{Dipartimento di Fisica, La Sapienza - Universit\`a di Roma , Piazzale Aldo Moro 2, 00185 Rome, Italy}

\author{Bernhardt L. Trout}
\email{trout@mit.edu}
\affiliation{Department of Chemical Engineering,
Massachusetts Institute of Technology,
77 Massachusetts Ave,
Cambridge MA 02139 USA}

\author{Leonardo Guidoni}
\email{leonardo.guidoni@univaq.it}
\affiliation{Dipartimento di Scienze Fisiche e Chimiche, Universit\`a degli studi de L'Aquila, Via Vetoio, 67100 Coppito, L'Aquila, Italy}

\begin{abstract}
The electronic properties of the oxygen molecule, in its singlet and triplet states, and of many small oxygen-containing radicals and anions have important roles in different fields of Chemistry, Biology and Atmospheric Science. Nevertheless, the electronic structure of such species is a challenge for ab-initio computational approaches because of the difficulties to correctly describe the statical and dynamical correlation effects in  presence of one or more unpaired electrons.
Only the highest-level quantum chemical approaches can yield reliable characterizations of their molecular properties, such as binding energies, equilibrium structures, molecular vibrations, charge distribution and polarizabilities.
In this work we use the variational Monte Carlo (VMC) and the lattice regularized Monte Carlo (LRDMC) methods to investigate the equilibrium geometries and molecular properties of oxygen and oxygen reactive species. Quantum Monte Carlo methods are used in combination with the Jastrow Antisymmetrized Geminal Power (JAGP) wave function ansatz, which has been recently shown to effectively describe the statical and dynamical correlation of different molecular systems. In particular we have studied the oxygen molecule, the superoxide anion, the nitric oxide radical and anion, the hydroxyl and hydroperoxyl radicals and their corresponding anions, and the hydrotrioxyl radical. 
Overall, the methodology was able to correctly describe the geometrical and electronic properties of these systems, through compact but fully-optimised basis sets and with a computational cost which scales as $N^3-N^4$, where $N$ is the number of electrons. This work is therefore opening the way to the accurate study of the energetics and of the reactivity of large and complex oxygen species by first principles.
\end{abstract}

\maketitle

\section{Introduction}


Oxygen molecule  and reactive oxygen species (ROS) are of great importance for several fundamental processes in Chemistry, Biology and Atmospheric research.
The dioxygen molecule itself, both in its  ground triplet state O$_2$ $X \, ^3 \Sigma_g^-$ 
and in its lowest excited singlet state
O$_2$ $a \, ^1 \Delta_g$,
has been extensively studied experimentally and theoretically, 
because it is involved in many natural photo-chemical and photo-biological processes such as photodegradation, aging,  photocarcinogenesis, etc.\cite{Schweitzer:2003p24197}
Moreover, several oxygen radicals, like hydroxyl radical,  superoxide anion and nitric oxide, are present under physiological conditions inside the cells, being involved in cell signaling, in redox  regulations \cite{Droge:2002p12375}, and in other processes involving cell damage, mutagenesis, cancer and degradation. They also are among the main actors of biological aging,\cite{Harman:1981} due to their  oxidative damage to DNA, proteins, lipids, and of other components of the cell.\cite{Beckman:1998wi}
Larger ROS, the polyoxides  and their radicals are also believed to be important for  atmospheric and environmental chemistry, chemistry of combustion and flames,  radiation chemistry,  and biochemical oxidations\cite{Cerkovnik:2013jv}. Other species, such as the hydroperoxyl radical and the hydrotrioxyl radical are also important reactive intermediates of interest in atmospheric chemistry.\cite{Murray:2009bf}

The molecular systems characterized by the presence of one or more unpaired electrons are often challenging for ab-initio computational quantum chemistry approaches, because they require a good description of both static and dynamical correlation.
These methods can be successfully applied to small and medium size reactive oxygen species, but their unfavorable scaling with the system size prevents the application to reactivity studies of larger complexes.
%
Recently, quantum Monte Carlo (QMC) approaches are emerging as a valuable and promising alternative to standard quantum chemical methods in the study of the electronic structure of molecules. Within QMC, the multi-dimensional integrals that arise in the calculation of physical observables or the application of projection operators are managed by stochastic methods. 
This intrinsic characteristic of the method has advantages and disadvantages. For instance, in Variational Monte Carlo, the expectation value of the Hamiltonian operator is minimized through the optimization of the variational parameters of a given wave function ansatz. The resulting estimated energy will be affected by a  stochastic error, that decreases as one over the square root of the computational time, and which is always much larger than the numerical errors of any standard deterministic approach. On the other side, QMC allows  to construct wave function ansatzes with a very large functional flexibility without adding cumbersome computational costs. For instance, in QMC wave functions it is typically introduced a bosonic terms (the Jastrow factor), with an explicit dependence on the inter-electronic distances.
The freedom in the definition of the wave function ansatzes that can incorporate electron correlation terms, as well as the stochastic nature of the QMC algorithms, translates in having accurate calculations that can be carried out on massively parallel high performance computers with a favorable scaling with the system size.
Recent applications of QMC methods on molecular systems were successful in describing high pressure hydrogen\cite{Attaccalite:2008p12639,Morales:2010p28600}, well depths of small molecular systems\cite{Toulouse:2008p27527,Braida:2011p27951}, excited states\cite{Zimmerman:2009hh,Guareschi:2013kt,Floris:2014cb}, transition metal complexes\cite{Casula:2009p12627}, binding energies of molecular or supramolecular complexes with noncovalent interactions\cite{Dubecky:2013fx,Ambrosetti:2014fz}, water molecule and water clusters \cite{Alfe:2013ks,Dagrada:2013uu,Sterpone:2008p12640,Zen:2013is}, diradical molecules\cite{Zen:2014dh}, and biomolecules \cite{Valsson:2010p25419,Coccia:2012ex,Filippi:2012hg,Coccia:2014do,Coccia:2014_new}.

In this work we used as wave-function ansatz the Jastrow Antisymmetrized Geminal Power\cite{Casula:2003p12694,Casula:2004p12689,Zen:2013is} (JAGP), which is based on the Pauling's resonating valence bond theory of the chemical bond. 
This anzatz has been used in several different contexts \cite{Sorella:2013wt,Sterpone:2008p12640,Coccia:2012fi,Barborini:2012iy,Stella:2011et,Marchi:2011p27883,Marchi:2009p12614,Beaudet:2008p12626,Sorella:2007p12646,Casula:2005p14146,Neuscamman:2013ej,Neuscamman:2013ku,Neuscamman:2012hm}, leading to significant advantages with respect to the Jastrow correlated single determinant ansatz (JSD), albeit having a comparable computational cost.
The Antisymmetrized Geminal Power (AGP) is indeed intrinsically multi-determinant, allowing the correct description of systems that cannot be described by JSD, as for instance shown in Ref.~\citenum{Zen:2014dh} for di-radicals.
A remarkable property of JAGP is that, in the limit of large Jastrow factor, it is size consistent for partitioning of the system in fragments of spin zero or spin $1/2$, as shown in Refs.~\citenum{Sorella:2007p12646,Neuscamman:2012hm}, and of course also 
when the total spin of the compound is equal to the sum of the spin of the fragments.
The computational cost of QMC calculations, both in the simplest variational Monte Carlo (VMC) level or in the more expensive lattice regularized diffusion Monte Carlo (LRDMC) level, for a JAGP wave function scales as $N^\gamma$, with $3<\gamma<4$ and $N$ the number of electrons in the system\cite{Coccia:2012kz}. Although there is a large prefactor in this scaling, QMC calculations uses algorithms which scale very efficiently up to hundreds of thousands processors, thus are feasible in the modern supercomputing facilities.
Several recent methodological developments on the QMC approaches have now made computationally affordable not only single point energy calculations, but also the evaluation of forces\cite{Sorella:2010p23644}, equilibrium structures\cite{Barborini:2012iy,Barborini:2012it,Coccia:2012kz,Coccia:2012ex,Coccia:2014do,Zen:2012br,Zen:2013is,Dagrada:2013uu}, 
vibrational properties\cite{Zen:2012br,Zen:2013is},  dipole, quadrupole, polarizability and other properties related to the electronic density\cite{Coccia:2012fi,Zen:2013is,Varsano:2014ef}.


A careful investigation of the performances of QMC methods based on JAGP on small but challenging reactive oxygen species, 
will allow us to assess the quality and versatility of this ansatz to correctly describe this systems. In this work we will evaluate several electronic and geometrical properties on neutral and ionic systems that were extensively studied in the literature such as triplet \cite{Bytautas:2005km,Medved:2001p23059,Poulsen:1998gm,Spelsberg:1994go,Kumar:1996bq,McDowell:1998gr} and
singlet dioxygen\cite{Schweitzer:2003p24197,Xu:2002ft},
the superoxide anion\cite{Ervin:2003bn,VanDuzor:2010p23154},
the hydroxyl and hydroperoxyl radicals\cite{Ruscic:2002p21517,Ajitha:1999p21418,Manohar:2006cr,Kallay:2003cc,Manohar:2007p21419,deGrey:2002ih,Lubic:1984hl,RienstraKiracofe:2002fb} and their anions,
the nitric oxide\cite{Medved:2001p23059,Polak:2004p23142,Huber:1979jv,BUNDGEN:1997ff}, and
the hydrotrioxyl radical\cite{Suma:2005jg,Mansergas:2007ew,Murray:2009bf,Varner:2009et,Denis:2009p23032,LePicard:2010jn,Anglada:2010kw,Varandas:2011ks,McCarthy:2012ja,Cerkovnik:2013jv,PabloAD:2008jg,Varandas:2011ks}.

The paper is organized as follows: 
in Section~\ref{sec:methods}
we shortly recall the main features of the QMC approaches and of the JAGP ansatz, reporting also the technical details of the calculations.
In Section~\ref{sec:results} our computational results will be reported and compared with other ab-initio approaches and experimental findings.
This section is conveniently divided in subsections, according to the system studied:
in Section~\ref{sec:res_atoms} we assess the capability of the VMC/JAGP and LRDMC/JAGP approach to correctly describe oxygen and nitrogen atoms;
in Section~\ref{sec:res_O2}  we discuss the triplet and singlet oxygen molecule and the superoxide anion; 
in Section~\ref{sec:res_OH}   the hydroxyl radical and anion;
in Section~\ref{sec:res_NO}   the nitric oxide radical and anion;
in Section~\ref{sec:res_HO2}   the hydroperoxyl radical and anion;
in Section~\ref{sec:res_HO3}   the hydrotrioxyl radical.
Our concluding remarks are finally reported in 
Section~\ref{sec:con}.
Moreover, in 
Appendix~\ref{app.scO2} we have reported the complete dissociation curve of the triplet oxygen molecule and discussed in detail the problem of size-consistency of the JAGP ansatz for this molecule.

\section{Methods and Computational Details}\label{sec:methods}

{\bf Quantum Monte Carlo techniques.} The wave function ansatz used in the QMC calculations presented in this paper is the Jastrow Antisymmetrized Geminal Power (JAGP),\cite{Casula:2003p12694,Casula:2004p12689,Zen:2013is} 
that is the product 
\begin{equation}\label{eq.JAGP}
\Psi_{JAGP}(\bar{\mathbf{x}}) = 
  \Psi_{AGP}(\bar{\mathbf{x}}) \cdot \Psi_J(\bar{\mathbf{x}}) 
\end{equation}
of the Antisymmetrized Geminal Power (AGP) function $\Psi_{AGP}$ and the Jastrow factor $\Psi_J$, where
$\bar{ \mathbf{x}}=(\bar{ \mathbf{r}},\bar{ \sigma})$  represents the collective electronic coordinates. 

The AGP is an antisymmetric function, that for an unpolarized system ({\em i.e.} zero total spin $S$) of $N=2 N_p$ electrons 
and $M$ atoms
is defined  as: 
\begin{eqnarray}\label{equ:AGP}
\Psi_{AGP}\left(\bar{\textbf{x}}\right) &=& 
  \hat{\cal A} 
     \prod_{i}^{N_p} G \left( \textbf{x}_{i};\textbf{x}_{N_p+i} \right) 
  , \\
G(\textbf{x}_{i} ;\textbf{x}_{j}) &=& 
   {\cal G}\left( \textbf{r}_{i},\textbf{r}_{j} \right) 
   \frac{ \delta_{\uparrow,\sigma_i} \delta_{\downarrow,\sigma_j} - \delta_{\downarrow,\sigma_i} \delta_{\uparrow,\sigma_j} }{\sqrt{2}} 
     \label{equ:calG}
   , \\
{\cal G}\left( \textbf{r}_{i},\textbf{r}_{j} \right) &=& 
  \sum_{a,b}^M \sum_{\mu_a} \sum_{\mu_b}  g^{a,b}_{\mu_a, \mu_b} \phi^a_{\mu_a}\left(\textbf{r}_{i}\right)\phi^b_{\mu_b}\left(\textbf{r}_{j} \right) ,
  \label{equ:calG}
\end{eqnarray} 
where $\hat {\cal A}$ is the antisymmetrization operator,  $G$ is the  geminal pairing function, $\cal G$ is its symmetric spatial part, $\phi^a_{\mu_a}$ ($\phi^b_{\mu_b}$) is the atomic orbital $\mu_a$ ($\mu_b$) of atom $a$ (b), with indexes $a$, $b$ running over all the M atoms and $\mu_a$ ($\mu_b$) over the orbitals. The elements  $g^{a,b}_{\mu_a, \mu_b}$ are wave function parameters.
The geminal $G$ is a spin singlet, so ${\cal G}$ has to be symmetric and $g^{a,b}_{\mu_a, \mu_b} = g^{b,a}_{\mu_b, \mu_a}$.
%
The AGP can be generalized to describe also polarized systems\cite{Casula:2003p12694} with total spin $S$:
let say that
the system is constituted by $N_p$ paired electrons and $N_u = 2 S$ unpaired electrons with same spin, that without loss of generality can be considered spin-up.
Thus, spin-up electrons are $N_{\uparrow} = (N_p + N_u)$, and spin-down electrons are $N_{\downarrow} = N_p$, for a total of $N=2 N_p +N_u$ electrons.
The generalized AGP wave function is defined as:
\begin{equation}
\Psi_{AGP}\left(\bar{\textbf{x}}\right) = 
   \hat{\cal A} \left\{ 
      \prod_{i}^{N_{p}} G\left( \textbf{x}_{i};\textbf{x}_{N_p+i} \right) 
      \prod_{j}^{N_u} {\Psi}_j \left( \textbf{x}_{2 N_p + j} \right) 
      \right\} ,
\label{eq:gagp1}
\end{equation}
where we have introduced $N_u$ single-electron functions 
$
{\Psi}_j ( \textbf{x}_i) = 
   \sum_a^M \sum_{\mu_a} \bar g^a_{\mu_a, j} \phi^a_{\mu_a} (\textbf{x}_i)  \delta_{\uparrow,\sigma_i} ,
$
that depends on the wave function parameters $\bar g^a_{\mu_a, j}$.

The Jastrow factor is a  symmetric positive function of the electronic positions that depends on the inter-particle  distances, and   describes the dynamical correlation among electrons and satisfies the electron-electron and electron-nucleus cusp conditions\cite{Foulkes:2001p19717,Drummond:2004p18505,Zen:2013is}. 
Since this term explicitly depends on the inter-electronic distances, the JAGP ansatz can proficiently and efficiently be used only within a quantum Monte Carlo\cite{Foulkes:2001p19717,Zen:2013is} approach.
More in detail, in our calculations we used the Jastrow factor 
$$J=\exp({U_{en} + U_{ee} + U_{een} + U_{eenn}}) , $$
that involves:
the one-electron interaction term $U_{en}$, 
the homogeneous two electron interaction term $U_{ee}$,
and the inhomogeneous two-electron interaction terms $U_{een}$ and $U_{eenn}$ (represently respectively an electron-electron-nucleus function  and 
an electron-electron-nucleus-nucleus function).
They are defined as follows:
\begin{eqnarray}
\label{eq.Jen}
U_{en}( \bar{\textbf{r}}) &=& 
  \sum_{a}^M \sum_{i}^N \left[
  - 
    Z_a { 1-e^{- b_1 \sqrt[4]{2 Z_a} r_{ia}} \over b_1 \sqrt[4]{2 Z_a} }
  + \sum_{\mu_a}
        f^a_{\mu_a} \chi^a_{\mu_a}(\textbf{r}_{ia}) 
      \right]
\\
\label{eq.Jee}
U_{ee}\left(\bar{\textbf{r}}\right) &=& 
  \sum_{i<j}^N \left[
    { 1-e^{- b_2 r_{ij}} \over 2 b_2 }
    \right]
\\
\label{eq.Jeen}
U_{een}\left( \bar{\textbf{r}} \right) &=& 
  \sum_{a}^M \sum_{i<j}^N \left[
      \sum_{\mu_a,\nu_a} 
        \bar f^{a}_{\mu_a,\nu_a} 
          \chi^a_{\mu_a}\left(\textbf{r}_{ia}\right)
          \chi^a_{\nu_a}\left(\textbf{r}_{ja}\right)
  \right] ,
\\
\label{eq.Jeenn}
U_{eenn}\left( \bar{\textbf{r}}\right) &=& 
  \sum_{a \neq b}^M \sum_{i<j}^N \left[
      \sum_{\mu_a} \sum_{\mu_b}
        \tilde f^{a,b}_{\mu_a,\mu_b} 
          \chi^a_{\mu_a}\left(\textbf{r}_{ia}\right)
          \chi^b_{\mu_b}\left(\textbf{r}_{jb}\right)
  \right] ,
\end{eqnarray}
where 
the vector 
$\textbf{r}_{ia}=\textbf{r}_i-\textbf{R}_a$
is the difference between the position of the nucleus $a$ and the electron $i$,  
$r_{ia}$ is the corresponding distance,
$r_{ij}$ is the distance between electrons $i$ and $j$,  
$Z_a$ is the electronic charge of the nucleus $a$,
$\chi^a_{\mu_a}$ are the atomic orbitals of nucleus $a$, 
and $b_1$, $b_2$, 
$f^a_{\mu_a}$, 
$\bar f^a_{\mu_a,\nu_a}$, 
$\tilde f^{a,b}_{\mu_a,\mu_b}$ 
are variational parameters.
The leading contribution for the description of electronic correlation is given by  $U_{ee}$, but also the inhomogeneous two-electron interaction terms $U_{een}$ and $U_{eenn}$ are particularly important in the JAGP ansatz, because they reduce the unphysical charge fluctuations included in the AGP function, as discussed in Refs.~\citenum{Sorella:2007p12646,Neuscamman:2012hm}.
%
Observe that the Jastrow factor is positive defined, thus the fermionic nature of electrons is taken into account only in the AGP term, that defines completely the nodal surface  of the JAGP ansatz ({\em i.e.} the hyper-surface on	which the wave function $\Psi_{JAGP}=0$, and	across which it changes sign). 
A more detailed description of the JAGP ansatz can be found in Refs.~\citenum{Casula:2003p12694,Zen:2013is,Zen:2014dh}.

In this work we report both variational and fixed node diffusion Monte Carlo results. As shown, our JAGP ansatz is functionally dependent on different  parameters which can be optimized, according the variational principle, to provide the lowest energy estimate within this given ansatz. 
The parameters of the wave functions for all the atoms and molecular systems reported in this work have been optimized by using the already validated and stable optimization schemes discussed in Ref.~\citenum{Zen:2013is}.
In particular, the optimization used for every  atom and molecule considered in its equilibrium configuration and without external field, starts from an initial configuration where the AGP matrix is diagonal, the exponents are initialized to values taken from standard Dunning's basis sets (where too small and too large values are eliminated because not necessary in the presence of our Jastrow factor, see discussion in Ref.~\cite{Zen:2013is}) and all the Jastrow parameters are set to zero, with the exception of $b_1=b_2=1$.
Next, the optimization procedure follows the protocol here reported:
{\em (i)} optimization of the AGP, namely of the matrix elements and the contraction coefficients of the basis set, with fixed  exponents and  Jastrow parameters $b_1=b_2=1$;
{\em (ii)} optimization of the AGP and relaxation of the values of the exponents of the AGP basis set and of the $b_1$ and $b_2$ parameters; 
{\em (iii)} optimization of the Jastrow terms, keeping the AGP parameters fixed; 
{\em (iv)} optimization of the overall JAGP, keeping fixed  the exponents in the basis set, both for the AGP and the Jastrow;
{\em (v)} optimization of all the parameters, including the exponents of the basis set, with increasing statistical accuracy. 
For the molecular systems considered in a configuration that is different from the equilibrium one, or in presence of an external field, we use a slightly different protocol: we take as initial configuration the one optimized at the equilibrium and with no external field, and we set to zero the parameters 
$f^a_{\mu_a}$, 
$\bar f^a_{\mu_a,\nu_a}$, 
$\tilde f^{a,b}_{\mu_a,\mu_b}$ 
of the Jastrow.
Next, we follow the optimization steps {\em (i)} to {\em (v)}, with the only difference that the values of the exponents in the AGP part of the wave function are kept fixed to the values of the optimized equilibrium configuration.

Molecular structures are also optimized, at the VMC level. In order to have a reliable structure, it is fundamental  to have accurate and efficiently calculated VMC forces. The implementation used here is based on the reweighting technique\cite{Zen:2013is} (to have forces with finite variance)  and on 
the Adjoint Algorithmic differentiation\cite{Sorella:2010p23644} (to compute all the components of the forces with a computational cost that does not grow as the number of atoms in the system).
For such optimization we have adopted a  steepest descent approach, already used successfully for several other molecular systems\cite{Barborini:2012iy,Barborini:2012it,Coccia:2012kz,Coccia:2012ex,Coccia:2014do,Zen:2014dh}.

The VMC results can further be improved by using the fixed-node projection Monte Carlo techniques, which provide the lowest possible energy with the constraint that the wave function $\Phi_{FN}$ has the same nodal surface of an appropriately chosen guiding function\cite{Reynolds:1982en,Foulkes:2001p19717}, which here is the variationally-optimized JAGP wave function $\Psi_{JAGP}$.
The fixed-node projection Monte Carlo method that we have adopted is the LRDMC \cite{Casula:2005p14138,Casula:2010p14082}, which is efficient also for systems with a large number of electrons\cite{Casula:2010p14082}
and  preserves the variational principle even when used in combination with nonlocal pseudopotentials\cite{Casula:2010p14082}.
All the reported  LRDMC results correspond to the continuous extrapolation (lattice mesh size $a\to 0$),  corresponding to the best variational results within the fixed node constraint given by $\Psi_{JAGP}$. 
Since the LRDMC calculations are much more demanding  than the VMC calculations, in terms of computational time, they have been carried out only in few crucial cases.

{\bf Computational details.}
The QMC calculations reported in this paper have been obtained using the {\em TurboRVB} package developed by S. Sorella and coworkers\cite{TurboRVB}, that includes a complete suite of variational and diffusion Monte Carlo codes for wave function and geometry optimization of molecules and solids.
The scalar-relativistic energy consistent pseudopotentials (ECP) of Burkatzki {\it et al.}\cite{Burkatzki:2007p25447} 
have been used in order to describe the two core electrons of the oxygen and nitrogen atoms, whereas the hydrogen atoms are described without  pseudopotential (the nuclear cusp is satisfied by the Jastrow factor).
As basis sets, in most of the calculations in the paper, we have used the hybrid contracted orbitals, introduced in Ref.~\citenum{Zen:2013is}, which allows to have a rapid convergence of the molecular properties with a relatively small number of variational parameters.
In more details, the  basis sets we have used 
for the AGP part are:
(9s,9p,3d,2f) contracted in \{12\} hybrid orbitals for the oxygen atom, 
(8s,7p,4d,3f) contracted in \{12\} hybrid orbitals for the nitrogen atom,
(7s,6p,2d)    contracted in \{4\}  hybrid orbitals for the hydrogen atom.
For each s, p, \ldots orbital type, there is one Slater type orbital (STO), introduced to improve the description of the diffusive part of the wave function, the remaining orbitals being of Gaussian type (GTOs). 
As basis sets for the atomic orbitals included in the inhomogeneous terms of the Jastrow factor, namely in $U_{en}$, $U_{een}$ and $U_{eenn}$ reported in Eqs.~\ref{eq.Jen},~\ref{eq.Jeen} and \ref{eq.Jeenn},
we used an uncontracted basis for the $U_{en}$ and $U_{een}$ term, and a contracted with hybrid orbitals basis for $U_{eenn}$.
This choice resulted very effective in having a converged basis set for the Jastrow while keeping the number of parameters of the wave function reasonably small.   
In more details, in $U_{en}$ and $U_{een}$  we used 
(5s,4p,2d,1f) basis set for the oxygen or nitrogen atoms, and 
(3s,2p,1d) for  hydrogen atom, whereas 
in  $U_{eenn}$ the orbitals are contracted in \{4\} hybrid orbitals for the oxygen or nitrogen atoms, and in  \{2\} hybrid orbitals for hydrogen atoms.
In some cases we have considered also other basis sets, in order to evaluate the basis set convergence. In these cases, the basis sets are described in the corresponding tables.

In the Results section we will use the acronyms VMC/JAGP/ECP and LRDMC/JAGP/ECP to indicate respectively the variational and the $a\to 0$ extrapolated lattice regularized diffusion Monte Carlo results, applied using the JAGP wave function ansatz with ECP pseudopotentials\cite{Burkatzki:2007p25447}.

{\bf Molecular properties.}
The dissociation energies of diatomic molecules ($ AB \to A + B $ ) has been calculated through the fitting of the potential energy surface nearby the minimum. We have firstly calculated the electronic energy  $E_{AB}(r)$ 
at some different values of the inter-atomic distance $r$ between $A$ and $B$.
We have therefore fitted the values of the energy difference
$ E_{AB}(r) - E_A - E_B $,
where $E_A$ and $E_B$ are the energies of the isolated atomic species
with a Morse potential:
\begin{equation}
f(r) = D_e  \left[ 1-e^{-a (r-r_e)} \right]^2 - D_e \, .
\label{eq:morse}
\end{equation}
The value of 
$D_e$ is the classical dissociation energy,  the values of 
$r_e$ is the classical equilibrium distance, and the force constant at the minimum is 
$k_e = 2 D_e a^2$.
The molecular vibrational energies for a Morse potential are:
\begin{equation}
G(n) = \omega_0 (n+1/2) - \frac{(\omega_0 (n+1/2))^2}{4 D_e}
\label{eq:ene_morse}
\end{equation}
where $n$ is the vibrational quantum number,  
$\omega_0 = \sqrt{k_e / m}$ is the harmonic frequency of vibration, and $m=(1/m_A+1/m_B)^{-1}$ is the reduced mass of $AB$.
From eq.~\ref{eq:ene_morse} we have estimated the fundamental frequency:
\begin{equation} 
\nu_0 = G(1)-G(0) = \omega_0 -{\omega_0^2 \over {2 D_e}} 
\label{eq:freq0}
\end{equation}
and the zero point energy (ZPE):
\begin{equation} 
\textrm{ZPE} = G(0) = {\omega_0 \over 2} - \frac{\omega_0^2}{ 16 D_e } \, .
\label{eq:ZPE}
\end{equation}
Thus, the dissociation energy $D_0$ at 0~K  has been finally estimated as 
$D_0 = D_e - ZPE$.
 
In the Results section we have also reported the ionization energy (IE), i.e. the energy difference between the neutral molecule and the corresponding cation, and the electron affinity (EA), i.e. the energy difference between the anion and the corresponding neutral molecule.
For the diatomic molecules we indicated with EA$_e$  the  adiabatic electron affinity without ZPE correction, and with EA$_0$ the adiabatic electron affinity at 0~K (i.e., considering also the $\Delta$ZPE).
 
In some cases we also reported the dipole moment $\mu$ and the polarizability $\alpha$, using the  estimators already introduced in Refs.~\citenum{Coccia:2012fi,Zen:2013is}.  In particular we calculated the polarizability by considering the dipole deviation induced by an external field of 0.01~au, which is in the linear response regime for all molecules, as we have verified by DFT calculations.

\section{Results}\label{sec:results}

\subsection{ Oxygen and Nitrogen atoms }\label{sec:res_atoms}

Before discussing the diatomic and polyatomic ROS, we firstly assessed the quality of the JAGP ansatz for the description of oxygen and nitrogen atoms, as show in Table~\ref{tab:atoms}. 
The experimental ionization energies and the electron affinities are compared to several computational methods: VMC/JAGP/ECP, LRDMC/JAGP/ECP, DFT/B3LYP/aug-cc-pVQZ, 
MP2/aug-cc-pVQZ and CCSD(T)/aug-cc-pVQZ.
The great accuracy of the QMC approach for the IE is clear also at the 
VMC/JAGP/ECP level, where both for oxygen and nitrogen are underestimated by $\sim0.14eV$ ($\sim1\%$ of the absolute value).
The use of LRDMC/JAGP/ECP slightly improves this estimation.
The EA of the oxygen is underestimated by 0.086~eV ($\sim6\%$) in VMC/JAGP/ECP, and of 0.048~eV ($\sim3\%$) in LRDMC/JAGP/ECP.
We also observe that the total energy of the LRDMC scheme improves the VMC energy of $\sim 0.01$H, both for nitrogen and oxygen atoms.

In summary, both the QMC approaches result of accuracy comparable or better than that obtained using MP2 and CCSD(T) calculations.
In particular, LRDMC/JAGP/ECP seems as accurate as CCSD(T), whereas VMC/JAGP/ECP is slightly less accurate, as expected.

\begin{table}\label{tab:atoms}
\caption{ Ionization Energy (IE) and Electron Affinity (EA) of nitrogen and oxygen atoms, calculated by VMC, LRDMC, DFT and other quantum chemical approaches versus experimental values. 
We also report the total energy of the QMC results for the JAGP/ECP function.
}
\begin{ruledtabular}
\begin{tabular}{ c l l  l l l  }
Atom & Method &  & Energy[H] & IE[eV] & EA[eV] \\
\hline
N \\
 & Experiment & Ref.~\citenum{cccbdb,RienstraKiracofe:2002fb}  && 14.534 & -0.07(2) \\

 & VMC/JAGP/ECP    & this work & -9.78464(2) & 14.388(1) &	-0.337(1) \\
 & LRDMC/JAGP/ECP  & this work & -9.79281(2) & 14.411(1) &	-0.151(8) \\

 & DFT/B3LYP/aug-cc-pVQZ & Ref.~\citenum{cccbdb}  && 11.929 & 2.962 \\
 & MP2/aug-cc-pVQZ & Ref.~\citenum{cccbdb}        && 14.619 & -0.688 \\
 & CCSD(T)/aug-cc-pVQZ & Ref.~\citenum{cccbdb}    && 14.500 & -0.230 \\

\hline
O \\
 & Experiment & Ref.~\citenum{cccbdb,RienstraKiracofe:2002fb}  && 13.618 & 1.462 \\

 & VMC/JAGP/ECP    & this work & -15.88428(3) & 13.482(1) &	1.376(1) \\
 & LRDMC/JAGP/ECP  & this work & -15.89500(5) & 13.520(1) &	1.414(1) \\

 & DFT/B3LYP/aug-cc-pVQZ & Ref.~\citenum{cccbdb}  && 14.142 & 1.682 \\
 & MP2/aug-cc-pVQZ       & Ref.~\citenum{cccbdb}  && 13.434 & 1.491 \\
 & CCSD(T)/aug-cc-pVQZ   & Ref.~\citenum{cccbdb}  && 13.514 & 1.399 \\

\hline
\end{tabular}
\end{ruledtabular}
\end{table}

\begin{table}
\caption{ Properties of the diatomic molecules and ions considered in the present work,
computed via QMC (VMC/JAGP/ECP and LRDMC/JAGP/ECP), are compared with experimental or estimated exact values.
We consider the zero temperature ($D_0$) and the classical ($D_e$) binding energy, the equilibrium bond length ($r_e$), the harmonic ($\omega_0$) and fundamental ($\nu_0$) vibrational frequencies, the zero point energy (ZPE), the classical (EA$_e$) and the zero temperature  (EA$_0$) electron affinity, the dipole ($\mu$) and the polarizability (
$\alpha_{\perp}$ 
and 
$\alpha_{\|}$ 
are respectively orthogonal and parallel to the molecular axes, $\left<\alpha\right>$ averaged). The stochastic error for the reported quantities evaluated by QMC are equal or lower than one unit in the last digit of the reported values, with the exception of the quantities in cm$^{-1}$, where the error is about 1\% of the reported value. }
\begin{ruledtabular}
\begin{tabular}{ l c  l l l l l l   l l  l  l l l }
Method & Ref. & $D_0$ & $D_e$ & $r_e$ & $\omega_0$ & $\nu_0$ & ZPE & EA$_0$ & EA$_e$ & $\mu$ & $\alpha_{\perp}$ & $\alpha_{\|}$ & $\left<\alpha\right>$ \\
       &  & [eV]  & [eV]  & [\AA]  & [cm$^{-1}$]  &  [cm$^{-1}$] &  [cm$^{-1}$] & [eV]  & [eV] & [Deb] & [au] & [au] & [au] \\
\hline
\multicolumn{7}{l}{ {\em triplet oxygen molecule:} O$_2$ $X \, ^3 \Sigma_g^-$  } \\
VMC   & this work & 4.72 &	4.82 &	1.196 &	1641 &	1606 &	816 & 0.39 &	0.36 && 7.65 &	15.63 &	10.31 \\
LRDMC & this work & & 5.02 &&&&&& 0.37 \\
Exp. & \citenum{Schweitzer:2003p24197,Bytautas:2005km,cccbdb,Ervin:2003bn,Olney:1997dc} & 5.12 & 5.22  & 1.207 & 1580 &  & 790 & 0.45 & 0.41 &&&& 10.54 \\
\hline 
\multicolumn{7}{l}{ {\em singlet oxygen molecule:} O$_2$ $a \, ^1 \Delta_g$  } \\
VMC   & this work & 3.64 &	3.73 &	1.203 &	1573 &	1532 &	781 & 1.48 &	1.45 \\
LRDMC & this work & & 3.84 &&&&&& 1.55 \\
Exp. & \citenum{Huber:1979jv,Schweitzer:2003p24197,cccbdb} & 4.14 &  & 1.215 & 1509 &  & 742 \\
\hline 
\multicolumn{7}{l}{ {\em superoxide anion:} O$_2^-$ $X \, ^2 \Pi_g$  } \\
VMC   & this work & 3.74 &	3.81 &	1.337 &	1173 &	1151 &	584 \\
LRDMC & this work & & 3.98 \\
Exp. & \citenum{Ervin:2003bn} & 4.10 &  & 1.348 & 1108 & 1090 & \\
\hline
\multicolumn{7}{l}{ {\em hydroxyl radical:} OH$^\bullet$ $X \, ^2 \Pi$  } \\
VMC   & this work & 4.35 &	4.58 &	0.965 &	3839 &	3640 &	1895 & 1.84 &	1.83 & 1.680 & 6.91 &	7.50 &	7.11 \\
LRDMC & this work &      & 4.62 &&&&&& 1.83 \\
Exp. & \citenum{Ruscic:2002p21517,cccbdb,NelsonJr:1967wb} & 4.41 & & 0.970 & 3738 && 1869 & 1.828 && 1.660\\
\hline 
\multicolumn{7}{l}{ {\em hydroxide anion:} OH$^-$ $X \, ^1 \Sigma$  } \\
VMC   & this work & 4.81 &	5.04 &	0.964 &	3795	 & 3618 &	1875 \\
LRDMC & this work & & 5.04 \\
Exp. & \citenum{cccbdb} & & & 0.964 & 3738 && 1869 \\
\hline
\multicolumn{7}{l}{ {\em nitric oxide radical:} NO$^\bullet$ $X \, ^2 \Pi$  } \\
VMC   & this work & 6.23 &	6.35 &	1.141 &	1973 &	1935 &	982 & -0.14 &	-0.18 & 0.138 & 9.58 &	16.41 &	11.86 \\
LRDMC & this work & & 6.44 &&&&&& -0.05 \\
Exp. & \citenum{cccbdb,Olney:1997dc,Polak:2004p23142,Huber:1979jv} & 6.48 & 6.61 & 1.151 & 1904 && 952 & 0.026  && 0.153 &&& 11.46 \\
\hline 
\multicolumn{7}{l}{ {\em nitric oxide anion:} NO$^-$ $X \, ^3 \Sigma$ } \\
VMC   & this work & 4.71 &	4.79 &	1.256 &	1410 &	1384 &	702 \\
LRDMC & this work & & 4.98 \\
Exp. & \citenum{cccbdb,Polak:2004p23142,Huber:1979jv} & & 5.14 & 1.258 & 1363 & \\
\hline
\end{tabular}
\end{ruledtabular}
\label{tab:diatomic}
\end{table}

\subsection{ Oxygen molecule and superoxide anion }\label{sec:res_O2}

\begin{figure}[htbp]
\caption{
Oxygen molecule dissociation. 
}\label{fig:O2}
\includegraphics[width=0.7\textwidth]{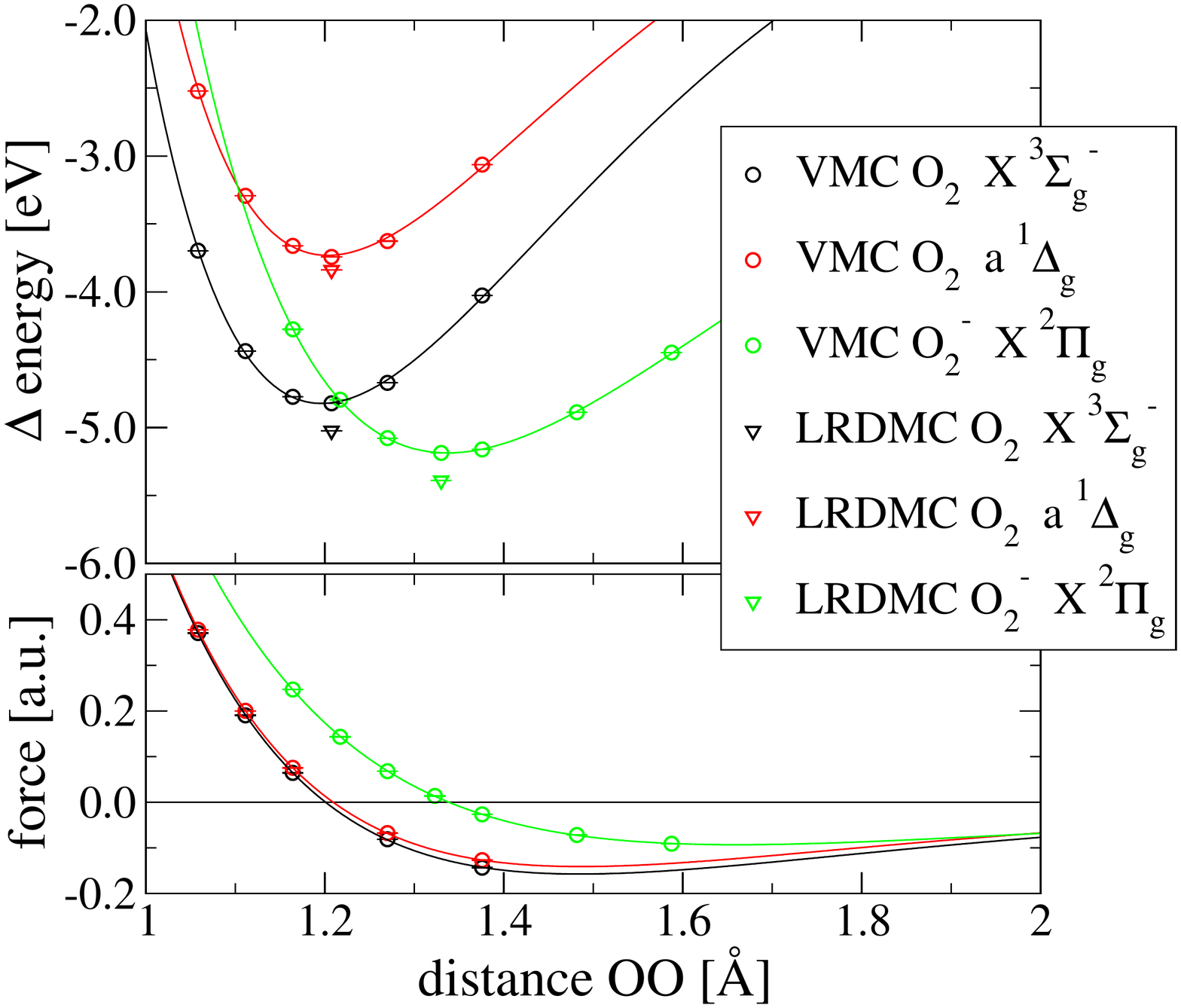}
\end{figure}

The quantitative evaluation of the energy gap between the triplet ground state and the first singlet excited state is a challenging for quantum chemistry approaches, and is estimated correctly only by the most accurate methods, as for instance CCSDTQ in Ref.~\citenum{Gadzhiev:2012dn}. Nevertheless, a correct evaluation of  this energy gap is fundamental for the  description of the reactivity of singlet oxygen, which is one of the most dangerous and reactive oxygen species. We have studied several properties of the 
triplet oxygen molecule O$_2$ $X \, ^3 \Sigma_g^-$ (the ground state),  
the
singlet oxygen molecule O$_2$ $a \, ^1 \Delta_g$ (the first exited state), 
and the
superoxide anion O$_2^-$ $X \, ^2 \Pi_g$.
In both spin states, for different values of the binding distance we have performed single point calculations of energies and forces, using VMC/JAGP/ECP approach as shown in Figure~\ref{fig:O2}.
By fitting the PES with a Morse potential, as described in Section~\ref{sec:methods}, 
we have evaluated  the bond length, the dissociation energy and the  vibrational properties.
The values are reported in  Table~\ref{tab:diatomic}.
Moreover, we have evaluated also the electron affinity, both for the triplet and the singlet oxygen, by considering the energy differences with the corresponding superoxide anions.
Finally, we have evaluated  the polarizability of the ground state triplet oxygen.
To increase the accuracy in the evaluation of the dissociation energy and the electron affinity, we also used LRDMC/JAGP/ECP single point calculations at the bonding length of each molecule.

To properly interpret the potential energy curves  we have to remind that 
the JAGP ansatz is not size consistent for O$_2$ dissociation (see the complete dissociation curve of the triplet oxygen molecule in the Appendix~\ref{app.scO2}), because the dissociated oxygen atoms have spin 1, whereas JAGP is size consistent only when the system dissociates in fragments having spin 0 or 1/2, or when the total spin of the compound is equal to the sum of the spin of the fragments \cite{Casula:2004p12689,Sorella:2007p12646}.
For this reason we must consider that the dissociation energy (calculated as the difference between the energy of the molecule and that of the isolated atoms)  is expected to be underestimated.
This is actually observed: for the O$_2$ $X \, ^3 \Sigma_g^-$ VMC underestimates both $D_0$ and $D_e$ of 0.40~eV ($\sim$10\% of the dissociation energy).
With the LRDMC approach one half of the missing dissociation energy is recovered,
indeed $D_e$ differs from the exact estimation only by 0.20~eV ($\sim$5\%).
For O$_2^-$ the VMC value of  $D_0$ is underestimated by 0.36~eV, value that  reduces to 0.19~eV for LRDMC, assuming that the ZPE correction is the same for VMC and LRDMC.
%
The electron affinity EA$_e$ of the triplet oxygen molecule is slightly underestimated at VMC (0.05~eV) and LRDMC (0.04~eV) level.

The VMC evaluations of the bond lengths, 
in the three considered cases, are all slightly underestimated ($\sim$0.01~\AA, corresponding to $\sim$1\% of the distance) with respect to experiments, but the overall agreement is good.
%
The vibrational frequency $\omega_0$, and consequently the fundamental frequency $\nu_0$ and the ZPE, are all overestimated:
+61~cm$^{-1}$ (3.9\%) the $\omega_0$ for the triplet oxygen,
+64~cm$^{-1}$ (4.2\%) the $\omega_0$ for the singlet oxygen, 
+65~cm$^{-1}$ (5.9\%) the $\omega_0$ for the superoxide.
In table~\ref{tab:O2_r-omega}
we report a comparison between the evaluations of $r_e$ and $\omega_0$ for the singlet and triplet oxygen molecule, obtained with VMC/JAGP/ECP calculations, with other computational methods, and with the experimental results. The accuracy of VMC/JAGP/ECP appears higher than MP2, MP4 and CCSD calculations, but lower than CCSD(T).
We also observe a coherence in the VMC errors for singlet and triplet oxygen evaluations: the bond lengths are underestimated by $\sim$1\%, and the frequency is overestimated by $\sim$4\%.

\begin{table}
\caption{ 
Equilibrium distance $r_e$ [\AA] and harmonic frequency of vibration $\omega_0$ [cm$^{-1}$]  of Triplet (O$_2\, X\,^3\Sigma$) and Singlet (O$_2\, a\,^1\Delta$) oxygen molecule, evaluated by several ab-initio methods, and compared with experimental value\cite{Huber:1979jv}. VMC results, obtained in this work, use the JAGP ansatz, ECP pseudopotential for the two core electrons, and the basis set described in Section~\ref{sec:methods}. Other results are taken from Ref.~\citenum{cccbdb}, and uses the aug-cc-pVQZ basis set, unless  explicitly indicated.
Errors [\%] are calculated in terms of deviation from the experimental value. 
}\label{tab:O2_r-omega}

\begin{ruledtabular}
\begin{tabular}{ l  c r c r  c r c r }
		& \multicolumn{4}{c}{ O$_2\, X\,^3\Sigma$ } & \multicolumn{4}{c}{	O$_2\, a\,^1\Delta$ } \\						
method	&	$r_e$	&	error	&	$\omega_0$ & error	&	$r_e$	&	error	&	$\omega_0$ & error		\\
\hline

HF		&	1.158	&	-4.1	\% &	1968	&	24.6	\% &	1.153	&	-5.1	\% &	1982	&	31.3	\% \\
LSDA		&	1.202	&	-0.4	\% &	1627	&	3.0	\% &	1.204$^a$ &	-0.9	\% &	1604$^a$ &	6.3	\% \\
BLYP		&	1.229	&	1.8	\% &	1493	&	-5.5	\% &	1.234$^b$ &	1.6	\% &	1465$^b$ &	-2.9	\% \\
B3LYP	&	1.204	&	-0.2	\% &	1637	&	3.6	\% &	1.203	&	-1.0	\% &	1626	&	7.8	\% \\
PBEPBE	&	1.218	&	0.9	\% &	1556	&	-1.5	\% &	1.221$^a$ &	0.5	\% &	1525$^a$ &	1.1	\% \\
MP2		&	1.219	&	1.0	\% &	1480	&	-6.3	\% &	1.243	&	2.3	\% &	1291	&	-14.4	\% \\
MP4		&	1.222	&	1.2	\% &	1464	&	-7.3	\% &	1.251$^a$ &	3.0	\% &	1257$^a$ &	-16.7	\% \\
CCSD		&	1.197	&	-0.8	\% &	1680	&	6.3	\% &	1.203	&	-1.0	\% &	1618	&	7.2	\% \\
CCSD(T)	&	1.208	&	0.1	\% &	1597	&	1.1	\% &	1.220	&	0.4	\% &	1498	&	-0.7	\% \\
VMC/ECP	&	1.196	&	-0.9	\% &	1641	&	3.9	\% &	1.203	&	-1.0	\% &	1573	&	4.2	\% \\
\hline
Experiment	&	1.207	&		&	1580	&		&	1.215	&		&	1509	&		\\
\hline
\multicolumn{9}{l}{ 
$^a$ basis set is aug-cc-pVTZ. 														
$^b$ basis set is cc-pVTZ.
}\\
\end{tabular}
\end{ruledtabular}
\end{table}

The VMC average polarizability $\left< \alpha \right>$ is only 0.23~au ($\sim$2\%) underestimated.  
Moreover, we can compare the parallel ($\alpha_{\|}^{VMC}$=15.63~au) and the perpendicular ($\alpha_{\perp}^{VMC}$=7.65~au) polarizability  with other very accurate quantum chemical calculations, and observe that the reported values are quite close to CCSD(T), 
$\alpha_{\|}^{CCSD(T)}$=15.03~au 
and 
$\alpha_{\perp}^{CCSD(T)}$=8.19~au 
from Ref.~\citenum{Medved:2001p23059},
or uncontracted MR-CI,
$\alpha_{\|}^{MR-CI}$=15.14~au
and
$\alpha_{\perp}^{MR-CI}$=7.88~au
from Ref.~\citenum{Spelsberg:1994go}.
These and others evaluations of the polarizability by several methods are reported in Table~\ref{tab:O2_polariz}, where it is  evident that the basis set convergence is fundamental for an accurate prediction of this property.
The proximity of our VMC result to the experiments and to the most accurate computational results provides an indication that our computational setup is able to catch the polarization properties even using a compact basis set with Slater type orbitals, as basis set D.
This fast convergence with the size of the basis set is due to the combined use of GTOs and STOs in the hybrid orbitals 
and to the fact that  the exponents of the gaussian primitives are variationally optimized within our QMC schemes; the latter point was already pointed out in ref. \cite{Coccia:2012fi}.

\begin{table}
\caption{ Polarizability of O$_2$ $X \, ^3 \Sigma_g^-$, calculated by several computational methods. For the VMC calculations we report also the energy and the variance. 
}\label{tab:O2_polariz}
\begin{ruledtabular}
\begin{tabular}{ l l  l l  l l l   }
Method &  & Ene. & Var. & $\alpha_{\perp}$ & $\alpha_{\|}$ & $\left<\alpha\right>$  \\
 & & [au] & [au] & [au] & [au] & [au] \\
\hline

B3LYP/cc-pVDZ 		& Ref.~\citenum{cccbdb} &&& 2.54 & 11.05 & 5.38 \\
B3LYP/cc-pVTZ 		& Ref.~\citenum{cccbdb} &&& 4.88 & 12.25 & 7.34 \\
B3LYP/cc-pVQZ 		& Ref.~\citenum{cccbdb} &&& 6.02 & 13.01 & 8.35 \\
B3LYP/aug-cc-pVDZ 	& Ref.~\citenum{cccbdb} &&& 7.86 & 14.07 & 9.93 \\
B3LYP/aug-cc-pVTZ 	& Ref.~\citenum{cccbdb} &&& 8.54 & 14.50 & 10.53\\
B3LYP/aug-cc-pVQZ 	& Ref.~\citenum{cccbdb} &&& 8.02 & 14.69 & 10.24\\

MP2/cc-pVDZ 			& Ref.~\citenum{cccbdb} &&& 3.16 & 8.01 & 4.78 \\
MP2/aug-cc-pVDZ 		& Ref.~\citenum{cccbdb} &&& 7.36 & 11.27 & 8.67 \\

CCSD(T)/aug-cc-pV5Z 	& Ref.~\citenum{Medved:2001p23059} &&& 8.19 & 15.03 & 10.47 \\
MR-CI 				& Ref.~\citenum{Spelsberg:1994go}  &&& 7.88 & 15.14 & 10.59 \\

VMC/JAGP/ECP/A$^a$  & this work & -31.93731(6) & 0.6680(7)  & 6.99(1) & 14.74(1) & 9.57(2) \\
VMC/JAGP/ECP/B$^b$  & this work & -31.93815(6) &	0.6572(15) & 7.17(1) & 15.20(1) & 9.85(1) \\
VMC/JAGP/ECP/C$^c$  & this work & -31.93966(6) & 0.5453(13) & 7.37(4) & 14.85(7) & 9.86(9) \\
VMC/JAGP/ECP/D$^d$  & this work & -31.94475(5) &	0.4961(7)  & 8.15(1) & 15.70(1) & 10.66(1) \\
VMC/JAGP/ECP/E$^e$  & this work & -31.94573(5) & 0.4771(4)  & 7.65(1) &	15.63(1) &	10.31(1) \\

\hline
Experiment & Ref.~\cite{Olney:1997dc} & &&&& 10.54 \\
Experiment & Refs.~\citenum{Kumar:1996bq,McDowell:1998gr} &&& 8.24 & 15.29 & 10.59 \\
\hline
\multicolumn{7}{l}{ $^a$ Basis set A is (6s,5p,3d)/\{8\} for the AGP, all GTOs, (4s,2p)/\{2\} for the Jastrow. }\\
\multicolumn{7}{l}{ $^b$ Basis set B is (6s,5p,3d)/\{8\} for the AGP, all GTOs, (4s,2p)/\{4\} for the Jastrow.   }\\
\multicolumn{7}{l}{ $^c$ Basis set C is (6s,5p,3d,2f)/\{8\} for the AGP, all GTOs, (4s,2p,1d)/\{2\} for the Jastrow. }\\
\multicolumn{7}{l}{ $^d$ Basis set D is (9s,9p,3d,2f)/\{8\} for the AGP, GTOs and STOs, (5s,4p,2d,1f)/\{4\} for the Jastrow,  see Section~\ref{sec:methods}.   }\\
\multicolumn{7}{p{17cm}}{ $^e$ Basis set E is (9s,9p,3d,2f)/\{12\} for the AGP, GTOs and STOs, (5s,4p,2d,1f)/\{4\} for the Jastrow,  see Section~\ref{sec:methods}. This basis is used for all the following calculations.   }\\
\end{tabular}
\end{ruledtabular}
\end{table}

In Figure~\ref{fig:density} we have reported a contour plot of the VMC/JAGP/ECP electronic density of 
O$_2$ $X \, ^3 \Sigma_g^-$, $a \, ^1 \Delta_g$, and O$_2^-$ $X \, ^2 \Pi_g$, each at the corresponding equilibrium distances.
It can be observed that the electron density distribution along the bond for the singlet and triplet oxygen molecule is similar, as expected for the fact that the two bond lengths are also similar. At opposite, in the case of the superoxide anion there is a lower electronic density at the middle of the bond, and a corresponding longer bond distance, despite the extra electron.

Finally, we have evaluated the triplet-singlet O$_2$ excitation energy , that is reported in Table~\ref{tab:O2_T-S}, in comparison with the experimental value and several others ab-initio evaluations.
Table~\ref{tab:O2_T-S} shows that all ab-initio methods overestimate the value of $\Delta E( ^1\Delta \leftarrow ^3\Sigma)$, of 137\% for HF, $>$50\% for DFT, $>$25\% for MP, \ldots
The overestimations can be rationalized considering that O$_2$ is a diradical, and the singlet state is more challenging to describe than the triplet ground state, in particular for single reference methods. 
\citet{Slipchenko:2002iz} have obtained an accurate estimation (error $<$10\%) by using spin-flip approach, and very recently 
\citet{Gadzhiev:2012dn} have obtained accurate results (error $\sim$5\%) 
by using single reference coupled clusters up to quadruple excitations.
The JAGP ansatz here used for our QMC evaluations should contain the leading ingredients to reliably describe singlet and triplet diradicals, as shown in Ref.~\citenum{Zen:2014dh} for the methylene and ethylene cases.
For the oxygen molecule our QMC evaluations of $\Delta E( ^1\Delta \leftarrow ^3\Sigma)$ have all an error $<$23\%, that makes that comparable, in terms of accuracy, to the CCSDT/cc-pVTZ evaluation of Ref.~\citenum{Gadzhiev:2012dn}.
However we observe, for the variational and fixed-node projection evaluations, the quite unusual behavior that larger basis sets and LRDMC calculations yields to estimations with a larger error, namely VMC with the  basis set C (see Table~\ref{tab:O2_T-S}) has an error of $\sim$2\%; VMC with largest basis has an error of $\sim$11\%; LRDMC with largest basis with $\sim$22\%.
This problem seems mainly due to the energies of the singlet O$_2$ with the two largest basis sets (E and F), that at the LRDMC level are unexpectedly $\sim$2mH higher than the LRDMC energy obtained with the smallest basis set (C).
However at the VMC level the largest basis sets have the lowest energies, indicating that there are no evident problems in the wave function optimizations, and the unusual, yet possible, behavior at the LRDMC level reflects the fact that variationally better wave functions at the variational level have not necessarily a better nodal surface.


\begin{table}
\caption{ 
Energies of Singlet (O$_2 a\,^1\Delta$) and Triplet (O$_2 X\,^3\Sigma$) oxygen molecule, and adiabatic energy separations $\Delta E( ^1\Delta \leftarrow ^3\Sigma)$, including the zero point energy correction $\Delta$ZPE,$^a$ computed by several ab-initio methods and compared with experimental value. Absolute energies of singlet and triplet oxygen are much lower for our QMC results than in other methods reported because the formers have pseudopotentials, the latter are all-electrons calculations. Stochastic error in QMC evaluations are $<10^{-4}$~H in the absolute energies, and are $<$0.003~eV for the value of $\Delta E$.
}\label{tab:O2_T-S}
\begin{ruledtabular}
\begin{tabular}{ l c  c c  c c r }
Method	&&	E(O$_2\,^3\Sigma$)	&	E(O$_2\,^1\Delta$)	&	$\Delta$ZPE	& \multicolumn{2}{c}{ $\Delta E( ^1\Delta \leftarrow ^3\Sigma)$ } \\
		&&	[H] 	&	[H]	&	[cm$^{-1}$]	&	[eV] & error	\\
\hline
HF/aug-cc-pVQZ	&	Ref.~\citenum{cccbdb}	&	-149.6931	&	-149.6077	&	-7	&	2.323	&	137	\%	\\
LSDA/aug-cc-pVTZ	&	Ref.~\citenum{cccbdb}	&	-149.6418	&	-149.5845	&	7	&	1.556	&	59	\%	\\
PBEPBE/aug-cc-pVTZ	&	Ref.~\citenum{cccbdb}	&	-150.2438	&	-150.1838	&	10	&	1.632	&	66	\%	\\
B3LYP/aug-cc-pVQZ	&	Ref.~\citenum{cccbdb}	&	-150.3957	&	-150.3345	&	5	&	1.666	&	70	\%	\\
MP4(full)/aug-cc-pVTZ	&	Ref.~\citenum{cccbdb}	&	-150.1742	&	-150.1281	&	92	&	1.242	&	27	\%	\\
MP2(full)/aug-cc-pVQZ	&	Ref.~\citenum{cccbdb}	&	-150.2200	&	-150.1746	&	93	&	1.224	&	25	\%	\\
QCISD/aug-cc-pVTZ	&	Ref.~\citenum{cccbdb}	&	-150.1235	&	-150.0711	&	32	&	1.423	&	45	\%	\\
QCISD(T)/aug-cc-pVTZ	&	Ref.~\citenum{cccbdb}	&	-150.1417	&	-150.0941	&	49	&	1.291	&	32	\%	\\
CCSD/cc-pVTZ	&	Refs.~\citenum{Gadzhiev:2012dn,cccbdb}	&	-150.1111	&	-150.0586	&	35	&	1.424	&	45	\%	\\
CCSD(full)/aug-cc-pVQZ	&	Ref.~\citenum{cccbdb}	&	-150.2191	&	-150.1669	&	31	&	1.415	&	44	\%	\\
CCSD(T)/cc-pVTZ	&	Refs.~\citenum{Gadzhiev:2012dn,cccbdb}	&	-150.1290	&	-150.0813	&	53	&	1.292	&	32	\%	\\
CCSD(T,full)/aug-cc-pVQZ	&	Ref.~\citenum{cccbdb}	&	-150.2400	&	-150.1928	&	49	&	1.277	&	30	\%	\\
CCSDT/cc-pVTZ	&	Ref.~\citenum{Gadzhiev:2012dn}	&	-150.1290	&	-150.0855	&		&	1.185	&	21	\%	\\
CCSDT(Q)/cc-pVTZ	&	Ref.~\citenum{Gadzhiev:2012dn}	&	-150.1309	&	-150.0931	&		&	1.029	&	5	\%	\\
CCSDTQ/cc-pVTZ	&	Ref.~\citenum{Gadzhiev:2012dn}	&	-150.1307	&	-150.0928	&		&	1.031	&	5	\%	\\
SF-CIS/cc-pVQZ	&	Ref.~\citenum{Slipchenko:2002iz}	&		&		&		&	1.447	&	48	\%	\\
SF-CIS(D)/cc-pVQZ	&	Ref.~\citenum{Slipchenko:2002iz}	&		&		&		&	1.067	&	9	\%	\\
SF-OF/cc-pVQZ	&	Ref.~\citenum{Slipchenko:2002iz}	&		&		&		&	1.061	&	8	\%	\\
VMC/ECP/C$^b$	&	this work	&	-31.9396	&	-31.9028	&	35	&	1.001	&	2	\%	\\
VMC/ECP/E$^b$	&	this work	&	-31.9457	&	-31.9061	&	35	&	1.078	&	10	\%	\\
VMC/ECP/F$^b$	&	this work	&	-31.9466	&	-31.9065	&	35	&	1.089	&	11	\%	\\
LRDMC/ECP/C$^b$	&	this work	&	-31.9735	&	-31.9332	&	35	&	1.098	&	12	\%	\\
LRDMC/ECP/E$^b$	&	this work	&	-31.9746	&	-31.9311	&	35	&	1.184	&	21	\%	\\
LRDMC/ECP/F$^b$	&	this work	&	-31.9748	&	-31.9308	&	35	&	1.197	&	22	\%	\\
\hline
Experiment	&	Ref.~\citenum{Huber:1979jv}	&		&		&		&	0.980	\\
\hline
\multicolumn{7}{l}{ $^a$ For VMC and LRDMC results, $\Delta$ZPE is estimated from for VMC/ECP/E; see 
Table~\ref{sec:res_O2}. }\\
\multicolumn{7}{l}{ $^b$ 
Basis set C is (6s,5p,3d,2f)/\{8\} for the AGP, all GTOs, (4s,2p,1d)/\{2\} for the Jastrow; 
} \\
\multicolumn{7}{l}{ 
basis sets E and F is (9s,9p,3d,2f)/\{12\} for the AGP, GTOs and STOs, } \\
\multicolumn{7}{l}{ 
and E uses (5s,4p,2d,1f)/\{4\} for the Jastrow,  
whereas F uses uncontracted (5s,4p,2d,1f). 
}\\						
\end{tabular}
\end{ruledtabular}
\end{table}

\subsection{ The hydroxyl radical OH$^\bullet$ and anion OH$^-$ }\label{sec:res_OH}

\begin{figure}[htbp]
\caption{
Hydroxyl radical and hydroxide anion dissociation. 
}\label{fig:OH}
\includegraphics[width=0.7\textwidth]{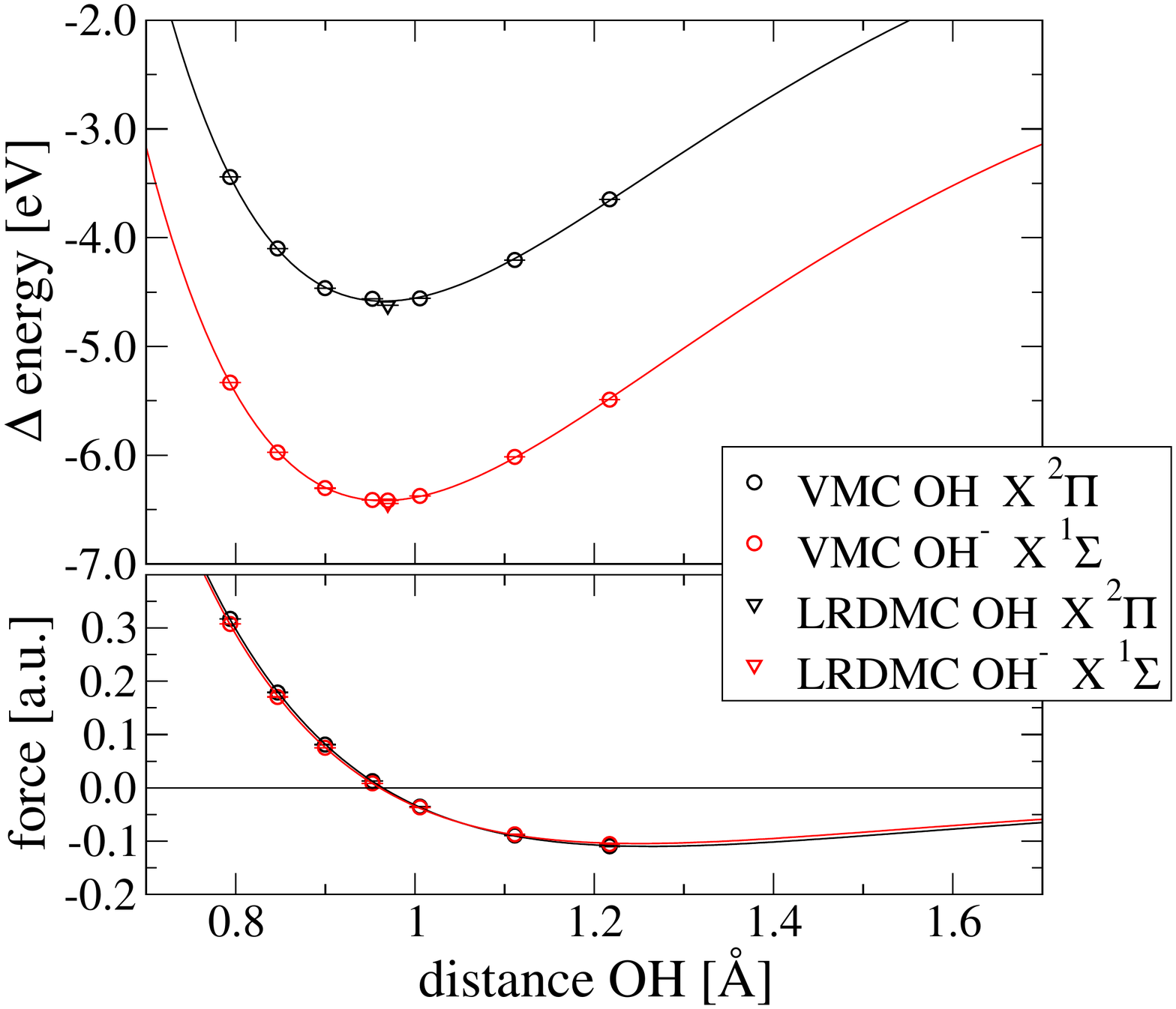}
\end{figure}

OH$^\bullet$ is an important radical that has been studied in several papers, in relation to its  molecular properties\cite{Ajitha:1999p21418,Ruscic:2002p21517,Kallay:2003cc,Manohar:2006cr,Manohar:2007p21419} and the reaction mechanism  with other molecules\cite{Zhou:2002p19041,Zhou:2005p23937,Mansergas:2006p23933,Ji:2010p19122,Bunkan:2013gi}.
Molecular properties of the  hydroxyl radical OH$^\bullet$ and its anion OH$^-$
were also computed  at VMC/JAGP/ECP and LRDMC/JAGP/ECP levels,  as reported in 
Table~\ref{tab:diatomic}.
The dissociation energies and forces are plotted in
Figure~\ref{fig:OH}.

As for the case of dioxygen, the JAGP ansatz is affected by a problem of size consistency when used to describe the  hydroxyl radical dissociation, because the dissociated oxygen atom fragment has spin 1.
However the hydrogen atom has spin 1/2, thus the problem should be mitigated with respect to the previously discussed O$_2$ case.
This is actually what can be observed in Table~\ref{tab:diatomic}.
The 0~K dissociation energy $D_0$ calculated by VMC is underestimated by only 0.06~eV ($\sim$1.5\%)  with respect to the experimental estimation, and this discrepancy reduces to 0.02~eV for LRDMC (assuming that the ZPE correction is the same for VMC and LRDMC).
The OH$^\bullet$ bond length calculated by VMC is only 0.005~\AA (0.5\%) shorter that the experimental estimation, whereas for OH$^-$ the agreement is perfect.
The vibrational frequency and the ZPE are slightly overestimated, as in the case of dioxygen:
+101~cm$^{-1}$ (2.7\%) the $\omega_0$ for OH$^\bullet$, 
 +57~cm$^{-1}$ (1.5\%) the $\omega_0$ for OH$^-$.
The calculated electron affinity is differing by the exact estimation only 0.01~eV.
The VMC/JAGP/ECP dipole is 
1.680~D,
very close to the exact estimation\cite{cccbdb,NelsonJr:1967wb} 
1.660~D,
and to other highly accurate quantum chemical approaches:
(see Table~\ref{tab:OH_polariz}).

In Table~\ref{tab:diatomic} we also report the VMC/JAGP/ECP polarizability of the hydroxyl radical, that is 
   $\alpha_\|^{VMC}$=7.50~au,
$\alpha_\perp^{VMC}$=6.91~au,
$\left< \alpha^{VMC}\right>$=7.11~au.
To the best of our knowledge, there are no experimental values that we can use as references, but we can compare our results with the evaluations obtained by other computational approaches, as shown in Table~\ref{tab:OH_polariz}.
All these results, and others available in Ref.~\citenum{cccbdb},
show a very strong dependence of the polarizability on the basis set.
However, as for the oxygen molecule, we are confident that our VMC results are close to the basis set convergence, thanks to the joint use of GTOs and STOs in the novel hybrid orbitals\cite{Zen:2013is}, and the fact that also the exponents are variationally optimized. 
In summary, our results demonstrate that VMC methods in combination with JAGP ansatz are able to accurately describe the molecular propertied of hydroxide radical and hydroxide anion.

\begin{table}
\caption{ Dipole [Deb] and Polarizability [au] of OH$^\bullet$ $X \, ^2 \Pi$, calculated with several computational methods.
}\label{tab:OH_polariz}
\begin{ruledtabular}
\begin{tabular}{ l l  l  l l l   }
Method &  & $\mu$ & $\alpha_{\perp}$ & $\alpha_{\|}$ & $\left<\alpha\right>$  \\
\hline

B3LYP/cc-pVDZ & Ref.~\citenum{cccbdb} & 1.673 & 2.36 & 6.04 & 3.59 \\
B3LYP/cc-pVTZ & Ref.~\citenum{cccbdb} & 1.688 & 4.07 & 7.16 & 5.10 \\
B3LYP/cc-pVQZ & Ref.~\citenum{cccbdb} & 1.677 & 4.63 & 7.78 & 5.68 \\
B3LYP/aug-cc-pVDZ & Ref.~\citenum{cccbdb} & 1.644 & 7.02 & 8.73 & 7.59 \\
B3LYP/aug-cc-pVTZ & Ref.~\citenum{cccbdb} & 1.650 & 7.61 & 8.84 & 8.02 \\
B3LYP/aug-cc-pVQZ & Ref.~\citenum{cccbdb} & 1.648 & 7.88 & 8.91 & 8.22 \\

MP2/cc-pVDZ & Ref.~\citenum{cccbdb} & 1.725 & 2.33 & 5.71 & 3.46 \\
MP2/aug-cc-pVDZ & Ref.~\citenum{cccbdb} && 6.72 & 8.33 & 7.26 \\

CVA-FSMRCC & Refs.~\citenum{Manohar:2006cr,Manohar:2007p21419} & 1.611 & & 6.61 & \\

finite field MRCCSD & Ref.~\citenum{Ajitha:1999p21418} & 1.627 \\
analytic MRCCSD     & Ref.~\citenum{Ajitha:1999p21418} & 1.650 \\
Full CI & Ref.~\citenum{Kallay:2003cc} & 1.685 \\

VMC/JAGP/ECP  & this work & 1.680 & 6.91 & 7.50 & 7.11 \\

\hline
Experiment & Refs.~\citenum{cccbdb,NelsonJr:1967wb} & 1.660 \\
\end{tabular}
\end{ruledtabular}
\end{table}

\subsection{ The nitric oxide radical NO$^\bullet$ and anion NO$^-$ }\label{sec:res_NO}

\begin{figure}[htbp]
\caption{
Nitric Oxyde dissociation. 
}\label{fig:NO}
\includegraphics[width=0.7\textwidth]{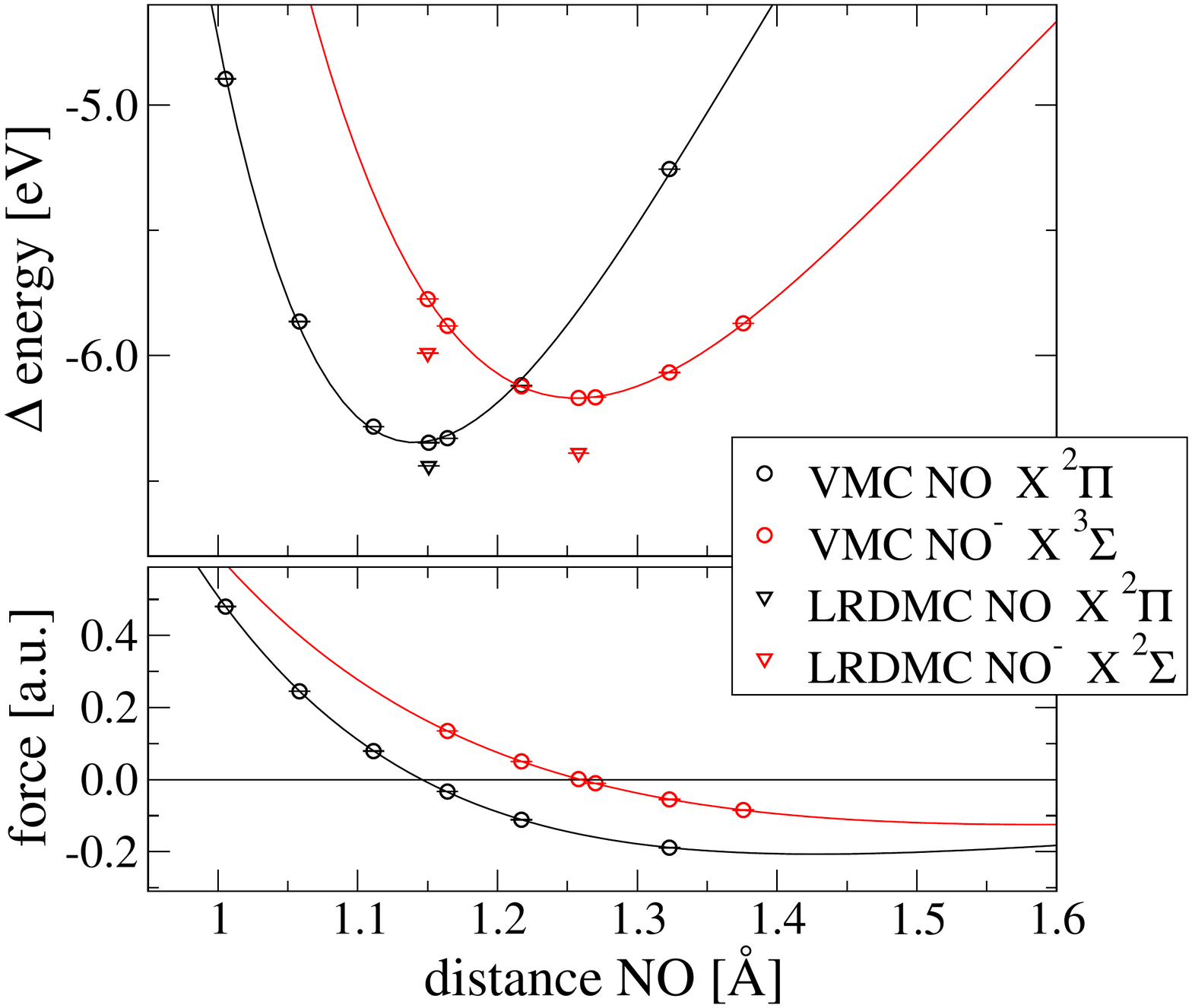}
\end{figure}

Nitric oxide (NO) is a reactive radical towards different molecules in the cell, such as thiols, oxygen-derived free radicals, and transition metal centers such as iron in heme.\cite{Hill:2010dg}  
Molecular simulations may help to unravel its complex role in cell signaling and redox properties, as well as the biochemical mechanisms that protect the cell from NO insults.\cite{Bocedi:2013ja}
The properties of the nitric oxide radical NO$^\bullet$ and anion NO$^-$ that we have calculated are reported in Table~\ref{tab:diatomic}, and the dissociation curves are reported in Figure~\ref{fig:NO}.

As for the previous cases, JAGP is affected by a problem of size consistency when used to study the nitric oxide radical, because the oxygen atom has spin 1 and the nitrogen atom spin 3/2.
Despite this problem, the binding energy $D_e$ of NO$^\bullet$ is underestimated only by 0.26~eV ($\sim$4\%) by VMC, a discrepancy which reduces to 0.17~eV ($\sim$3\%) for LRDMC; whereas for NO$^-$ $D_e$ is underestimated by 0.35~eV ($\sim$7\%) using VMC and 0.16~eV ($\sim$3\%) using LRDMC.
The VMC evaluation of the bond length of NO$^\bullet$ is underestimated by 0.010~\AA~ ($\sim$1\%), whereas for NO$^-$ it is almost exact.
Also in this case, the vibrational frequencies and the ZPE are overestimated with respect to the experimental values:
+69~cm$^{-1}$ (3.6\%) the $\omega_0$ of NO$^\bullet$, 
+47~cm$^{-1}$ (3.3\%) the $\omega_0$ of NO$^-$.

It can be observed in Figure~\ref{fig:NO} that the nitric oxide radical and anion reach the minimum of the PES at two different distances, but VMC/JAGP/ECP predicts that the minimum of the NO$^-$ is higher in energy than the minimum of NO$^\bullet$ of 0.18~eV. 
This yields to a negative electronic affinity EA$_e^{VMC}$=-0.18~eV, see Table~\ref{tab:diatomic}, that reduces to EA$_0^{VMC}$=-0.14~eV by including  the ZPE correction.
The LRDMC approach shifts the evaluation towards zero:
EA$_e^{LRDMC}$=-0.05~eV, 
EA$_0^{LRDMC}$=-0.01~eV (assuming that the ZPE correction is the same of the VMC case). 
These results are not very far from the almost null, albeit positive, electron affinity (0.026~eV) observed in experiments.

\begin{table}
\caption{ Dipole [Debye] and Polarizability [au] of NO$^\bullet$ $X \, ^2 \Pi$, calculated with several computational methods.
}\label{tab:NO_polariz}
\begin{ruledtabular}
\begin{tabular}{ l l  l  l l l   }
Method &  & $\mu$ & $\alpha_{\perp}$ & $\alpha_{\|}$ & $\left<\alpha\right>$  \\
\hline

B3LYP/cc-pVDZ & Ref.~\citenum{cccbdb} & 0.074 & 4.57 & 11.73 & 6.95 \\
B3LYP/cc-pVTZ & Ref.~\citenum{cccbdb} & 0.108 & 6.46 & 12.97 & 8.67 \\
B3LYP/cc-pVQZ & Ref.~\citenum{cccbdb} & 0.120 & 7.84 & 13.81 & 9.83 \\
B3LYP/aug-cc-pVDZ & Ref.~\citenum{cccbdb} & 0.117 & 9.62 & 15.12 & 11.45 \\
B3LYP/aug-cc-pVTZ & Ref.~\citenum{cccbdb} & 0.140 & 9.20 & 15.29 & 11.23 \\
B3LYP/aug-cc-pVQZ & Ref.~\citenum{cccbdb} & 0.142 & 9.36 & 15.38 & 11.37 \\

CVA-FSMRCCSD & Ref.~\citenum{Manohar:2007p21419} & 0.186 & 9.96 & 14.72 & 11.19 \\

MP2/cc-pVDZ & Ref.~\citenum{cccbdb} & 0.120 & 4.35 & 8.38 & 5.70 \\
MP2/aug-cc-pVDZ & Ref.~\citenum{cccbdb} && 9.35 & 11.48 & 10.06 \\

CCSD(T)/aug-cc-pV5Z & Ref.~\citenum{Medved:2001p23059} &  & 9.23 & 15.07 &  11.43 \\

VMC/JAGP/ECP  & this work & 0.138 & 9.58 & 16.41 & 11.86 \\

\hline
Experiment & Ref.~\citenum{Olney:1997dc} & 0.153 & && 11.46 \\
Experiment & Refs.~\citenum{McDowell:1998gr,BUNDGEN:1997ff} && 9.67 & 15.24 & 11.53 \\

\end{tabular}
\end{ruledtabular}
\end{table}

The molecular dipole and polarizabilities evaluated at VMC level results quite close to the experimental values.
In table~\ref{tab:NO_polariz} we can compare the VMC results with other computational approaches.
Similarly to the previous cases, the basis set convergence seems very important for this kind of calculations, and the VMC evaluation seems very accurate.

\subsection{ The hydroperoxyl radical HOO$^\bullet$  and anion HOO$^-$ }\label{sec:res_HO2}

HOO$^\bullet$ in a radical molecule of biological and biomedical importance\cite{deGrey:2002ih}, that has been extensively studied also by several ab-initio approaches\cite{Liskow:1971kq,Ajitha:1999p21418,Manohar:2006cr,Manohar:2007p21419}, in relation to its molecular and electronic properties. 
The geometrical parameters of the  hydroperoxyl radical HOO$^\bullet$  and anion HOO$^-$ have been optimized by VMC/JAGP/ECP.
The resulting values are reported in Table~\ref{tab:OOH}, and compared with the experimental values of Ref.~\citenum{Lubic:1984hl} and other computational approaches.
The VMC evaluations are very close to the experimental geometrical parameters for HOO$^\bullet$: both the OO  and the OH distances are only 0.005~\AA~ underestimated, and the OOH angle differs only by 0.1 degrees.
The comparison with other computational methods, Table~\ref{tab:OOH}, shows that this order of agreement is excellent.
We also reported the dipole $\mu$ of the HOO$^\bullet$, that is in agreement with previous MRCCSD calculations.\cite{Ajitha:1999p21418}

\begin{table}
\caption{ Properties of hydroperoxyl radical OOH$^\bullet$ and anion OOH$^-$. 
}\label{tab:OOH}
\begin{ruledtabular}
\begin{tabular}{ l l  c  l l l  l  l l  }
Molecule & Method && r(OO) & r(OH) & $\theta$(OOH) & $\mu$ & EA$_e$ & EA$_0$   \\
                 &&& [\AA] & [\AA] & [deg]         & [Deb] & [eV]   & [eV]     \\
\hline
OOH$^\bullet$ \\
	& Experiment & Ref.~\citenum{Lubic:1984hl,RienstraKiracofe:2002fb} & 1.3305(8) & 0.971(2) & 104.3(3) &&& 1.078(6) \\

    & VMC/JAGP/ECP   & this work                    & 1.3250(1) & 0.9661(1) & 104.43(2) & 2.1286(4) & 1.051(2) \\
    & LRDMC/JAGP/ECP & this work                                                                &&&&& 1.075(3) \\

        & HF/aug-cc-pVQZ    & Ref.~\citenum{cccbdb} & 1.2981 & 0.9458 & 106.307 & 1.976 & -0.588 & -0.557 \\
        & B3LYP/aug-cc-pVQZ & Ref.~\citenum{cccbdb} & 1.3251 & 0.9751 & 105.542 & 2.201 & 0.991 & 1.020 \\
        & MP2/aug-cc-pVTZ   & Ref.~\citenum{cccbdb} & 1.3132 & 0.9740 & 104.702 &  & 1.263 & 1.293 \\
        & QCISD(T)/cc-pVTZ  & Ref.~\citenum{cccbdb} & 1.3390 & 0.9716 & 103.770 && 13.194 & 13.226 \\
        & CCSD(T)/cc-pVTZ   & Ref.~\citenum{cccbdb} & 1.3360 & 0.9714 & 103.903 && 0.211 & 0.245 \\
        & MRCCSD & Ref.~\citenum{Ajitha:1999p21418} &&&& 2.130 & \\
\hline
OOH$^-$	\\
	& VMC/JAGP/ECP & this work & 1.5125(2) & 0.9538(2) &  98.48(3) \\

        & HF/aug-cc-pVQZ    & Ref.~\citenum{cccbdb} & 1.4538 & 0.9370 & 102.186 \\
        & B3LYP/aug-cc-pVQZ & Ref.~\citenum{cccbdb} & 1.5089 & 0.9606 & 99.358 \\
        & MP2/aug-cc-pVTZ   & Ref.~\citenum{cccbdb} & 1.4964 & 0.9625 & 97.762 \\
        & QCISD(T)/cc-pVTZ  & Ref.~\citenum{cccbdb} & 1.5313 & 0.9580 & 95.985 \\
        & CCSD(T)/cc-pVTZ   & Ref.~\citenum{cccbdb} & 1.5303 & 0.9579 & 96.069 \\
        &CCSD(T)/aug-cc-pVTZ& Ref.~\citenum{cccbdb} & 1.5285 & 0.9623 & 97.495 \\
\hline
\end{tabular}
\end{ruledtabular}
\end{table}

VMC and LRDMC energy calculations for the radical and its anion,  in their VMC optimized structures, 
yield an electron affinity EA$_e$ of 1.051(2)~eV for VMC, and 1.075(3)~eV for LRDMC.
These values are very close to the experimental\cite{RienstraKiracofe:2002fb} value of EA$_0$, 1.078(6)~eV, although for a fair comparison between the values the ZPE correction should be considered.
We did not calculated the VMC vibrational properties of HOO$^\bullet$  and HOO$^-$ in this work, but we can assume from other approaches, see Table~\ref{tab:OOH}, that they are of the order of 0.03~eV, 
so that a corrected value should be of about  EA$_0$  $\sim$1.05~eV.
Thus, the VMC and LRDMC calculations of the electronic affinity appears the most accurate between all the calculations reported in Table~\ref{tab:OOH}. The most problematic point of the other computational approaches appears to be the choice of a  small basis set, a problem which doesn't effect our fully optimized variational wave functions.

In figure~\ref{fig:density}
we have represented the electronic density of   OOH$^\bullet$ and  OOH$^-$, as obtained by VMC calculations.
It can be observed that the density of OOH$^\bullet$ along the OO bond is more similar to that of the superoxide anion 
than to that of the oxygen molecule. 
This observation reflects the fact  that the length of the OO bond of OOH$^\bullet$ (VMC: 1.325~\AA) is closer to the bond length of O$_2^-$ (VMC: 1.337~\AA) than  to the one of O$_2$ (VMC: 1.196~\AA).
In OOH$^-$ the electronic density along the OO bond is smaller in the middle and the length increase to 1.512~\AA.

\subsection{ The hydrotrioxyl radical HOOO$^\bullet$ }\label{sec:res_HO3}

\begin{figure}[htbp]
\caption{
Trans and cis  isomers of hydrotrioxyl radical (HOOO$^\bullet$). 
}\label{fig:HOOO}
\includegraphics[width=0.5\textwidth]{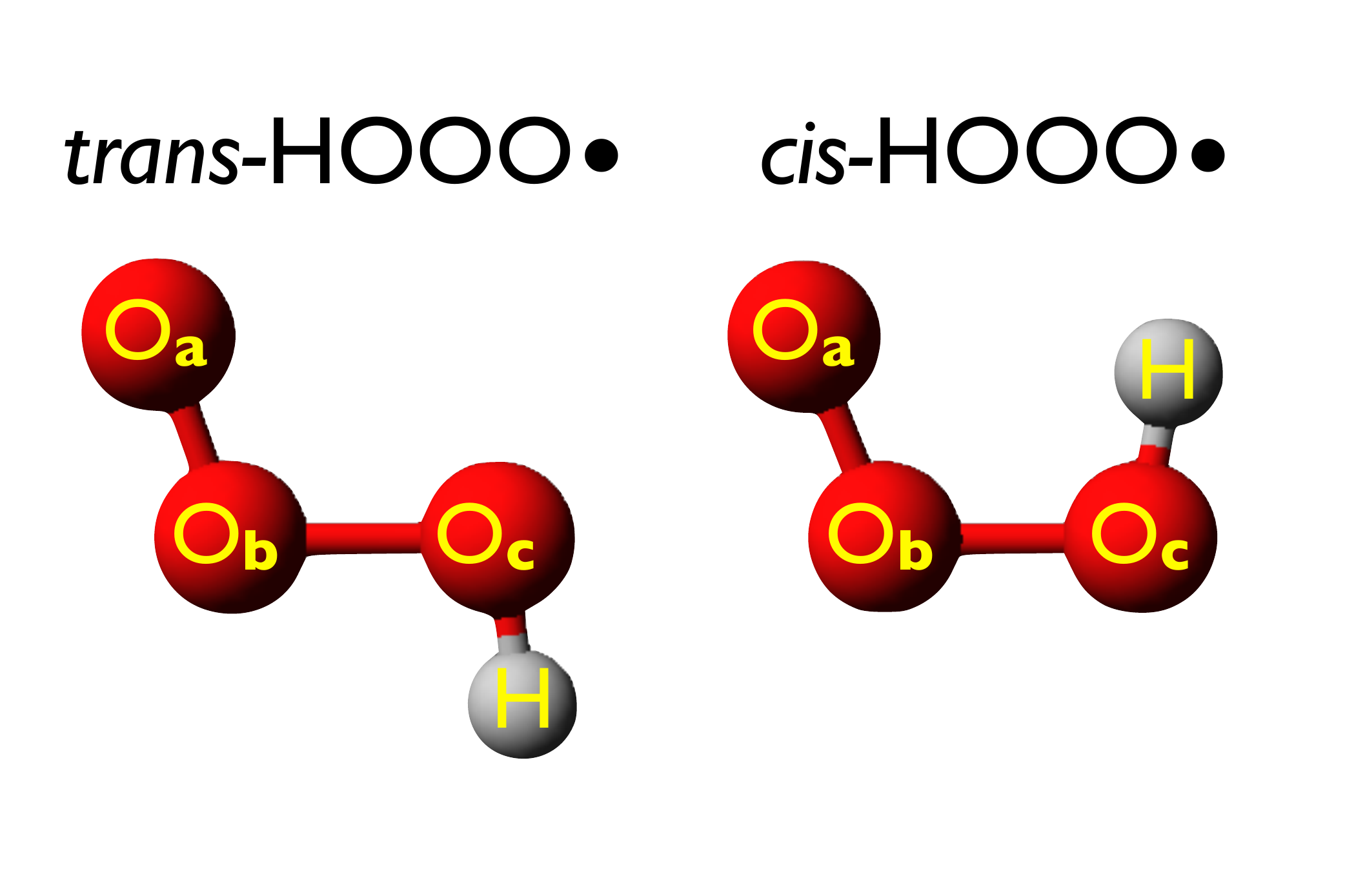}
\end{figure}

\begin{figure}[htbp]
\caption{
VMC electron density distributions for the triplet and singlet O$_2$ molecule and the superoxide O$_2^-$, the hydroperoxyl radical HOO$^\bullet$ and anoin HOO$^-$, the trans and cis isomers of the hydrotrioxyl radical, trans-HOOO$^\bullet$ and cis-HOOO$^\bullet$. 
The reported isosurfaces, that are cut in proximity of the plane of the molecule,  correspond to a value of the density of 0.0001, 0.001, 0.01 (white), 0.05 (silver), 0.1 (gray), 0.2 (green), 0.4 (yellow), 0.6 (orange) a.u-. 
}\label{fig:density}
\includegraphics[width=1.0\textwidth]{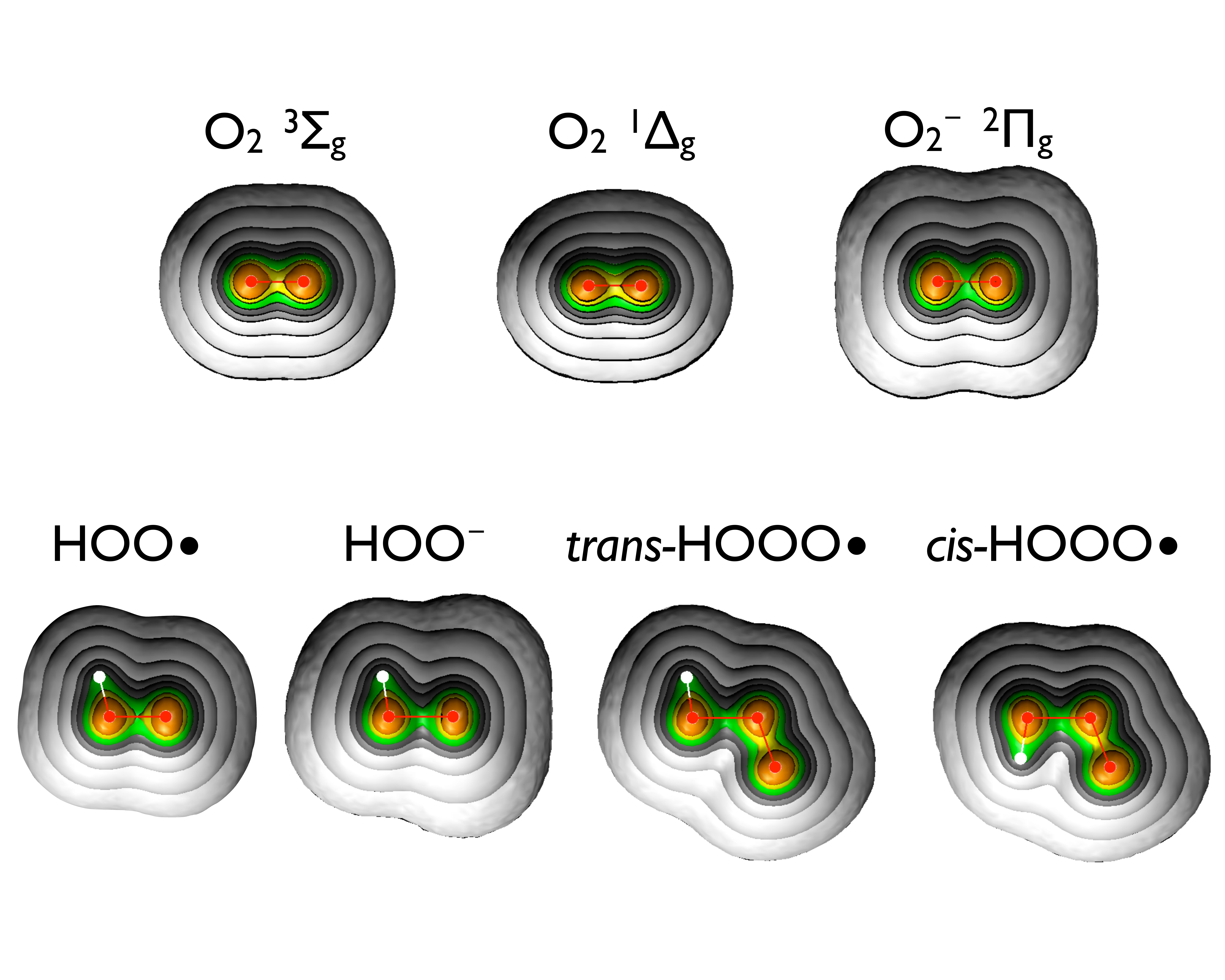}
\end{figure}

The HOOO$^\bullet$ radical is relevant in atmospheric processes since it is an intermediate in reactions involving the hydroxyl radical and molecular oxygen.\cite{Murray:2009bf}
For this reason, it has been extensively studied both experimentally\cite{Suma:2005jg,Derro:2008ey,LePicard:2010jn} and computationally\cite{Mansergas:2007ew,Varandas:2011ks,PabloAD:2008jg,Anglada:2010kw}.
VMC/JAGP/ECP structural optimizations have been performed for the hydrotrioxyl radical HOOO$^\bullet$, both for the cis and trans isomers, see figure~\ref{fig:HOOO}.
The geometrical parameters are reported in Table~\ref{tab:OOOH}.
The VMC dipole moment in the VMC optimized configurations are:
$\mu_{trans}=$2.0811(5)~D
and 
$\mu_{cis}=$1.1282(5)~D.
In figure~\ref{fig:density} we represent the electronic density of the cis and trans molecules.

\begin{table}
\caption{ Equilibrium structure of  trans and cis  isomers of hydrotrioxyl radical, see fig.~\ref{fig:HOOO}, optimized with VMC/JAGP/ECP, 
in comparison with experiments and other computational methods. Both structures are planar.
}\label{tab:OOOH}
\begin{ruledtabular}
\begin{tabular}{ l  l c  l l l  l l   }
 & Method$^b$ && r(O$_a$O$_b$) & r(O$_b$O$_c$) & r(O$_c$H) & $\theta$(O$_a$O$_b$O$_c$) & $\theta$(O$_b$O$_c$H) \\ 
&&& [\AA] & [\AA] & [\AA] & [deg] & [deg] \\ 
\hline
{\em trans}
\\
& Experiment & Ref.~\citenum{Suma:2005jg} & 1.225 & 1.688 & 0.972 & 111.02 & 90.04 \\ 

& VMC/JAGP/ECP & this work & 1.2318(2) & 1.5310(4) & 0.9649(1) & 109.78(4) & 98.32(2) \\ 

& QCISD/B2 				& Ref.~\citenum{Mansergas:2007ew} & 1.244 & 1.522 & 0.967 & 109.1 & 98.4 \\ 
& CCSD(T)/AVTZ				& Ref.~\citenum{Mansergas:2007ew} & 1.251 & 1.513 & 0.972 & 109.2 & 98.4 \\ 

& CCSD(T)-C6/AVQZ					& Ref.~\citenum{Varandas:2011ks} & 1.2248 & 1.5887 & 0.9693 & 109.70 & 97.41 \\ 

& UCCSD(T)/AVQZ				& Ref.~\citenum{PabloAD:2008jg} & 1.2265 & 1.5911 & 0.9694 & 109.76 & 97.35 \\ 

& CASSCF(19,15)/B2		& Ref.~\citenum{Mansergas:2007ew} & 1.223 & 1.758 & 0.976 & 110.8 & 94.2 \\ 

& 	CASSCF(19,15)/AVTZ		& Ref.~\citenum{Anglada:2010kw} & 1.226 & 1.674 & 0.973 & 109.9 & 96.9 \\ 

& MRCI/B1					& Ref.~\citenum{Mansergas:2007ew} & 1.233 & 1.647 & 0.960 & 107.4 & 96.6 \\ 
& CASPT2 (13,11)/B2		& Ref.~\citenum{Mansergas:2007ew} & 1.211 & 1.739 & 0.972 & 110.5 & 94.9 \\ 

& 	CASPT2(13,11)/AVTZ		& Ref.~\citenum{Anglada:2010kw} & 1.214 & 1.734 & 0.973 & 110.7 & 95.2 \\ 
& 	CASPT2(19,15)/AVTZ		& Ref.~\citenum{Anglada:2010kw} & 1.221 & 1.682 & 0.971 & 110.2 & 95.8 \\ 

& MRCI-C6/VTZ			& Ref.~\citenum{Varandas:2011ks} & 1.2203 & 1.6951 & 0.9684 & 110.48 & 95.03 \\ 

& MRCI+Q/AVTZ & Ref.~\citenum{Suma:2005jg} & 1.225 & 1.677 & 0.972 & 110.2 & 95.9 \\ 

\hline
{\em cis}
\\
& VMC/JAGP/ECP & this work$^a$ & 1.2503(1) & 1.4821(8) & 0.9680(2) & 111.80(4) & 99.59(2) \\ 

& CCSD(T)-C6/AVQZ					& Ref.~\citenum{Varandas:2011ks} & 1.2445 & 1.5289 &  0.9723 & 111.77 & 97.71 \\ 

& UCCSD(T)/AVQZ				& Ref.~\citenum{PabloAD:2008jg} & 1.2481 & 1.5265 & 0.9724 & 111.76 & 97.74 \\ 
& MRCI-C6/VTZ			& Ref.~\citenum{Varandas:2011ks} & 1.2443 & 1.5805 &  0.9709 & 111.96 & 96.21 \\ 

\hline
\multicolumn{8}{l}{$^b$ VxZ and AVxZ stand respectively for cc-pVxZ and aug-cc-pVxZ; B1 for 6-311+G(d,p); B2 for 6-311+G(2df,2p). }\\
\end{tabular}
\end{ruledtabular}
\end{table}

The only available experimental structure\cite{Suma:2005jg} is for the  {\em trans}-HOOO$^\bullet$.
As shown by table ~\ref{tab:OOOH} the geometrical parameters reported using different approaches are more scattered than what observed for previous molecules. In particular the bond length of the central OO bond is quite sensitive to both the method and the basis set.
In this case we  observe that the VMC optimized structure is not so close to the experimental value as in the molecules studied previously. 
The configurational parameter of {\em trans}-HOOO$^\bullet$ that mostly differs between VMC/JAGP/ECP and the experimental one is the length of the central OO bond, r(O$_b$O$_c$)=1.5310(4)~\AA~ for VMC, that underestimates the experimental value of 0.157~\AA~ by 9.3\%, whereas r(O$_a$O$_b$) is overestimated by 0.007~\AA, 
r(O$_c$H) is underestimated by 0.007~\AA, 
the angle $\theta$(O$_a$O$_b$O$_c$) is underestimated by 1.24~deg,
and $\theta$(O$_b$O$_c$H) is overestimated of 8.28~deg.
Thus,  r(O$_b$O$_c$) has an error that is one order of magnitude larger that what usually observed for VMC.
On the other hand, the description of the equilibrium geometry of this molecule is particularly challenging for all the ab-initio computational approaches; 
CCSD(T) and other single-reference methods also underestimate r(O$_b$O$_c$) of $\sim$0.1\AA, 
and results close to experimental value are obtained only  using expensive multi-reference methods\cite{Suma:2005jg,Varandas:2011ks} or active space approaches with very large active spaces, such as   CASPT2(19,15)/aug-cc-pVTZ in Ref.~\citenum{Anglada:2010kw}.

For the cis structure, VMC predicts a distance r(O$_b$O$_c$) of 1.4821(8)~\AA, that is 0.049~\AA~ shorter than the trans configuration.
This behavior of having a smaller r(O$_b$O$_c$) for the cis is in agreement with other computational calculation, see CCSD(T) and MRCI in table~\ref{tab:OOOH}.

Figure~\ref{fig:density} shows that the electronic density along this bond is very similar in the cis and trans configurations, and also to the HOO$^-$.
The value of $r(O_aO_b)$ is, both for cis and trans, slightly larger than the distance for the oxygen molecule, 
and the value of $r(O_cH)$ is almost the same of that of OH$^\bullet$.

The HOOO$^\bullet$ dissociates in OH$^\bullet$ (X $^2 \Pi$) and O$_2$ (X $^3 \Sigma_g^-$).
Thus, also in this case the JAGP ansatz is affected by a problem of size consistency, that could be responsible for the underestimation of the length of the O$_a$O$_b$ bond.
We expect that this problem would also influence the evaluation of the 
dissociation energy,
especially for this highly challenging system, where it is known, both from experiments\cite{Derro:2008ey,LePicard:2010jn} and highly accurate quantum chemical calculations\cite{Anglada:2010kw}, to be of the order of a few kcal/mol.
%
The classical dissociation energy $D_e$(HO-OO) (no ZPE correction) that we obtain by VMC/JAGP/ECP for the trans isomer is  
-1.77(5)~kcal/mol,
while for the cis isomer is
-1.35(5)~kcal/mol.
Both the results have the wrong sign, so the evaluation is also qualitatively wrong.
If we perform an LRDMC calculation on the VMC optimized structures, we obtain a $D_e$(HO-OO) of 
0.41(8)~kcal/mol
for the trans isomer and 
0.81(9)~kcal/mol
for the cis isomer.
These results shows that LRDMC calculations improve the dissociation energy, which has a positive sign.
Diffusion calculation can therefore alleviate the size consistency issue, although still underestimating the value of $D_e$ $\sim$5.8~kcal/mol obtained by \citet{Anglada:2010kw} using CASPT2(19,15) calculations,  which is compatible with the experimental data on the trans conformer\cite{LePicard:2010jn}.

\section{Conclusions}\label{sec:con}

In this paper we have calculated through QMC methods several chemical properties of small reactive oxygen species of interest for biological and atmospheric sciences.
Despite their small size, these compounds, which includes many radicals, are quite challenging for ab-initio computational approaches, because they are characterized by a strong electronic correlation, that is adequately described only by very accurate approaches.
The QMC evaluations here obtained have been compared with those of other computational methods and with experimental values.
We have demonstrated that the JAGP ansatz is able to provide a reliable description of these systems: VMC evaluations are often more accurate than CCSD ones with large basis sets, and the computationally most expensive LRDMC approaches often further improve  the results.
Our JAGP ansatz takes a great advantage from the use of the atomic hybrid orbitals introduced in Ref.~\citenum{Zen:2013is}, which allow us to have an almost converged basis set with a reasonably small number of fully optimized variational parameters in the wave function.
These converged basis sets lead to  very accurate QMC estimations of molecular dipoles and polarizabilities.

The major element that still slightly biases the QMC results is represented by the fact that JAGP is not size-consistent for partitioning of the system in fragments of spin higher than 1/2 (unless the total spin of the compound is equal to the sum of the spin the fragments). This leads in some cases to the underestimation (up to about 10\%) of the VMC molecular binding energies, which is often alleviated by LRDMC calculations. Further research efforts to improve the JAGP ansatz and solve this problem are under progress.
One possibility is to recover the size consistency by breaking the spin symmetry in the wave function; another is to
further improve  the description of static correlations by including triplet pairing correlations in the AGP part, namely by means of the  Pfaffian wave function.\cite{Bajdich:2006p18510,Bajdich:2008p18507}

Despite these small discrepancies, the VMC/JAGP level of theory provides us  the possibility to have geometrical and electronic properties of reactive oxygen species with a reasonable computational cost. Most notably, VMC calculations have a more favorable scaling with the system size than traditional post Hartree-Fock methods. This implies that with this accurate methodology, larger reactive oxygen species are affordable, and, most importantly,  their reactivity with other molecules in biological and atmospheric sciences can be investigated.

\appendix

\section{ The size-consistency problem of JAGP for the dissociation of the triplet O$_2$  molecule } \label{app.scO2}

\begin{figure}[htbp]
\caption{
The large plot reports the dissociation curve for the triplet (in black and blue respectively for VMC and LRDMC) and quintuplet (in red and orange respectively for VMC and LRDMC) O$_2$ molecule.
The green curve represents a Morse function, with the parameters set to the experimental values, see Table~\ref{tab:diatomic}. 
The inset reports the error of the VMC and LRDMC results for the O$_2$ triplet, calculated as the difference with the Morse function.
The three lowest plots represents the results obtained at the equilibrium and at a distance of 0.847~\AA, and of 3.704~\AA, for the LRDMC calculations with lattice mesh $a$ of 0.2, 0.3, 0.4, 0.5, and extrapolation $a \to 0$.  
}\label{fig:O2dissociation}
\includegraphics[width=0.7\textwidth]{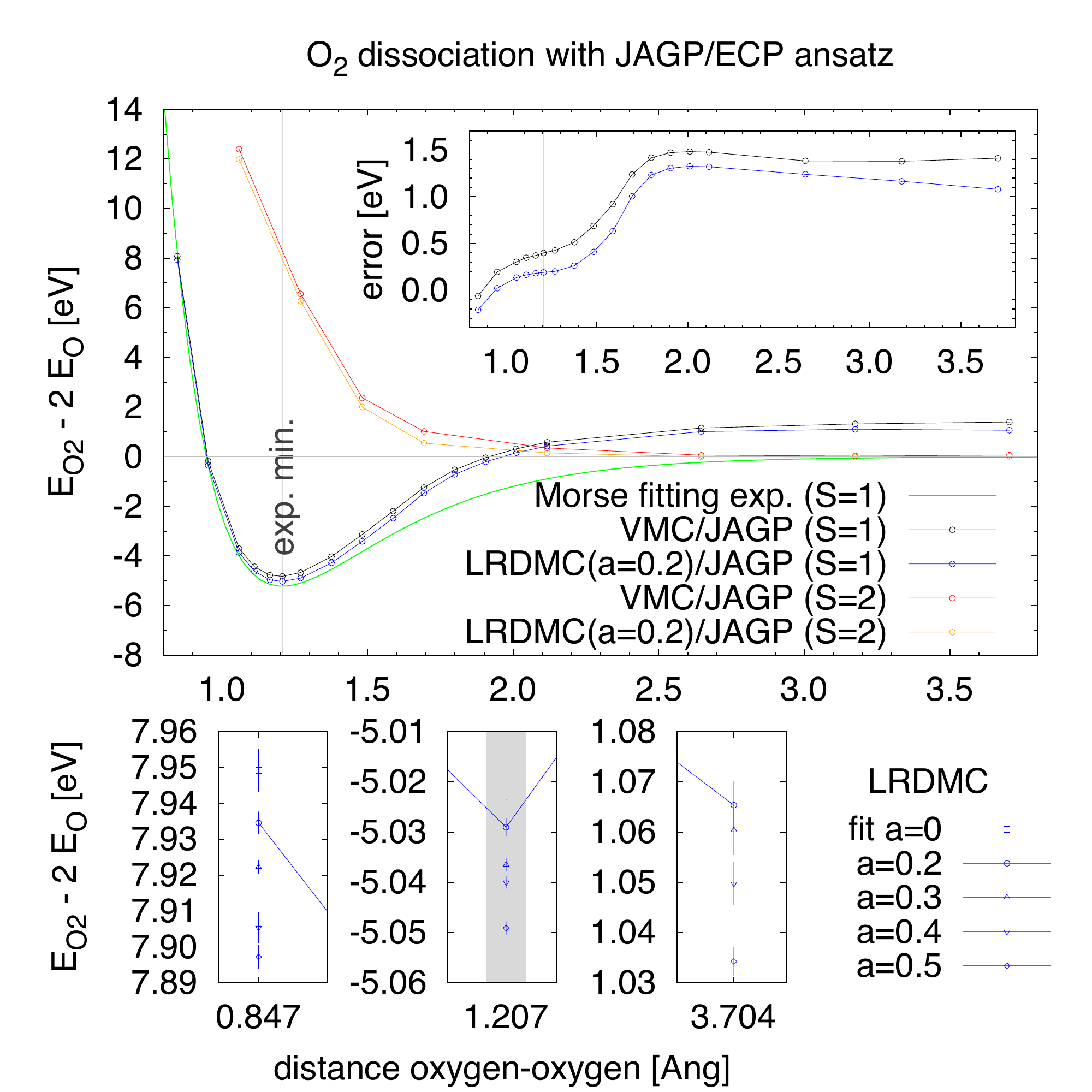}
\end{figure} 

As mentioned in the paper, the JAGP ansatz has a problem of size-consistency for the triplet O$_2$ molecule (total spin S=1, {\em i.e.} 2 unpaired electrons), because the oxygen atom has also a triplet ground state, and with the JAGP ansatz it is not possible to correctly describe the molecule in the dissociation regime, where the overall system of S=1 is factorized in two subsystems of S=1.
In this appendix, see fig.~\ref{fig:O2dissociation}, we report and discuss  the dissociation curve of the triplet oxygen molecule, investigate using the JAGP ansatz and via both the VMC and LRDMC schemes, in order to quantify the size of the error that is encountered in this case.
The JAGP results have been compared with a Morse function (green curve in fig.~\ref{fig:O2dissociation}) which has the parameters set according to the experimental values (see Table~\ref{tab:diatomic}).
As a further comparison, we have also reported the dissociation curve for an O$_2$ system with total spin S=2 ({\em i.e.} 4 unpaired electrons), that is a case where JAGP is size-consistent, as evident in the figure, both at the VMC (red curve) and LRDMC (orange curve).
The plot shows clearly (see the inset in fig.~\ref{fig:O2dissociation}) that the triplet O$_2$ dissociates with a slope than is sharper that Morse functions, as a consequence of the problem of size-consistency of JAGP in this system. The VMC curve is only $\sim$0.4~eV above the Morse function at the equilibrium distance (1.207~\AA), but the difference rapidly increase to $\sim$1.5~eV in at a distance of 1.9~\AA, and remains almost constant for larger distances. Thus, if we consider the energy difference at equilibrium and at large distance as an estimation of the binding energy, this would be $\sim$6.2~eV, that is overestimated, differently from the estimation adopted previously in the paper (anyway, we prefer the estimation $E_{O_2}-2 E_O$, because the electronic structure of the triplet O$_2$ at large distances is unreliable due to the problem of size-consistency).
The LRDMC curve (in blue) slightly improves the VMC curve, but overall also the LRDMC results have large deviations in the stretched region. At the equilibrium distance, the LRDMC is above the experimental curve by $\sim$0.2~eV, and the energy difference sharply increases up to $\sim$1.3~eV at an oxygen-oxygen distance of 1.9~\AA, before decreasing to $\sim$1.0~eV at 3.7~\AA. This is an indication that the lack of size-consistency  induces a problem in the nodal surface,  preventing the fixed-node projection scheme to improve significantly the VMC results.
The LRDMC curve reported is calculated with a lattice mesh of $a=0.2~a.u.$.
At the equilibrium distance and in the leftmost and rightmost points of the curve we have also evaluated the LRDMC energy for $a$ equal to 0.3, 0.4, 0.5, and the extrapolation to $a\to 0$, as shown in the lowest three plots in the figure.
These results show that the use of $a=0.2$ induces a bias in the curve that is lower than $\sim$0.01~eV. 
The inset in fig.~\ref{fig:O2dissociation} shows an asymmetry in the error around the equilibrium, that is small and rapidly converging to zero at distances smaller than that of equilibrium, and it is large and rapidly increasing at distances larger than the equilibrium.

\section*{Acknowledgement}
The European Research Council Project MultiscaleChemBio (n. 240624) within the VII Framework Program of the European Union has supported this work.
We thank the Singapore-MIT Alliance for funding.
The authors acknowledge Prof. Sandro Sorella for the assistance in the use of the TurboRVB Quantum Monte Carlo code and Matteo Barborini for critical reading of the paper and for valuable discussions.
Computational resources were provided by the PRACE Consortium, CINECA Computing Center and the Caliban-HPC Lab of the University of L'Aquila. 



\end{document}